\documentclass{aa}  

\usepackage{graphicx}
\usepackage{txfonts}
\usepackage{layouts}
\usepackage{enumerate}
\usepackage{natbib}
\usepackage{color}
\usepackage{amssymb}
\usepackage{amsmath}
\usepackage{rotating}
\usepackage[dvipsnames]{xcolor}
\begin{document}

   \title{Gaia eclipsing binary and multiple systems. \\Supervised classification and self-organizing maps}

   \titlerunning{{\it Gaia} Eclipsing Binary and Multiple Systems}

   \author{M.~S\"uveges \inst{1,2}
          \and F.~Barblan \inst{3}
          \and I.~Lecoeur-Ta\"ibi\inst{1}
          \and A.~Pr\v sa\inst{4}
          \and B.~Holl\inst{3}
          \and L.~Eyer\inst{3}
          \and A.~Kochoska\inst{5}
          \and N.~Mowlavi\inst{3}
          \and L.~Rimoldini\inst{1}
          }

   \institute{Dept.~of Astronomy, University of Geneva,
              Chemin d'Ecogia 16, CH-1290 Versoix, Switzerland \\
              \email{Maria.Suveges@unige.ch}
   \and
             Max-Planck-Institut f\"ur Astronomie, K\"onigstuhl 17, 69117 Heidelberg, Germany (present address) \\ 
             \email{sueveges@mpia-hd.mpg.de}
   \and
             Dept.~of Astronomy, University of Geneva,
             Chemin des Maillettes 51, CH-1290 Versoix, Switzerland
   \and
             Villanova University, Dept.~of Astrophysics and Planetary Science, 800 Lancaster Ave, Villanova PA 19085, USA
   \and
             Dept.~of Physics, University of Ljubljana, Jadranska 19, SI-1000 Ljubljana, Slovenia
}

   \date{Received ; accepted }
 
  \abstract
    {Large surveys producing tera- and petabyte-scale databases require machine-learning and knowledge discovery methods to deal with the overwhelming quantity of data and the difficulties of extracting concise, meaningful information with reliable assessment of its uncertainty. This study investigates the potential of a few machine-learning methods for the automated analysis of eclipsing binaries in the data of such surveys.}
   {We aim to aid the extraction of samples of eclipsing binaries from such databases and to provide basic information about the objects. We intend to estimate class labels according to two different, well-known classification systems, one based on the light curve morphology (EA/EB/EW classes) and the other based on the physical characteristics of the binary system (system morphology classes; detached through overcontact systems). Furthermore, we explore low-dimensional surfaces along which the light curves of eclipsing binaries are concentrated, and consider their use in the characterization of the binary systems and in the exploration of biases of the full unknown {\it Gaia} data with respect to the training sets.}
{ We have explored the performance of principal component analysis (PCA), linear discriminant analysis (LDA), random forest classification and self-organizing maps (SOM) for the above aims. We pre-processed the photometric time series by combining a double Gaussian profile fit and a constrained smoothing spline, in order to de-noise and interpolate the observed light curves. We achieved further denoising, and selected the most important variability elements from the light curves using PCA. Supervised classification was performed using random forest and LDA based on the PC decomposition, while SOM gives a continuous 2-dimensional manifold of the light curves arranged by a few important features. We estimated the uncertainty of the supervised methods due to the specific finite training set using ensembles of models constructed on randomized training sets.}
   {We obtain excellent results (about 5\% global error rate) with classification into light curve morphology classes on the {\it Hipparcos} data. The classification into system morphology classes using the Catalog and Atlas of Eclipsing binaries ({\it CALEB}) has a higher error rate (about 10.5\%), most importantly due to the (sometimes strong) similarity of the photometric light curves originating from physically different systems. When trained on {\it CALEB} and then applied to {\it Kepler}-detected eclipsing binaries subsampled according to {\it Gaia} observing times, LDA and SOM provide tractable, easy-to-visualize subspaces of the full (functional) space of light curves that summarize the most important phenomenological elements of the individual light curves. The sequence of light curves  ordered by their first linear discriminant coefficient is compared to results obtained using local linear embedding. The SOM method proves able to find a 2-dimensional embedded surface in the space of the light curves which separates the system morphology classes in its different regions, and also identifies a few other phenomena, such as the asymmetry of the light curves due to spots, eccentric systems, and systems with a single eclipse. Furthermore, when data from other surveys are projected to the same SOM surface, the resulting map yields a good overview of the general biases and distortions due to differences in time sampling or population.}
   {}


   \maketitle
%

\section{Introduction}

The advent of large-scale surveys in astronomy has opened new horizons in the investigation of our Galaxy and the Universe, thanks to the amount of information contained in the data \citep[perhaps the earliest, best-known example is the Sloan Digital Sky Survey;][]{yorketal00,eisensteinetal11,alametal15}. However, the sheer size of these databases requires the application of new types of analysis: those aimed precisely at the discovery of previously unknown knowledge in huge data sets through automated methods. Human interaction can be present only in a few crucial stages, and can deal with only subsamples of very limited size. Detailed assessment of each individual result or careful manual engineering of the procedures for each individual object are impossible, even though such big samples necessarily bring up many extremes, rare or exceptional cases. Investigations without the input of preliminary knowledge, letting the data alone determine the model, can also provide insight from a new angle into unsolved or less-frequently investigated issues. In the framework of astronomical databases, even an everyday task like the extraction of an appropriate sample for a specific research requires the database to be organized and easily accessible, and therefore the data to be pre-processed and some fundamental selective information extracted and made available in the database.
Accordingly, the use of machine-learning technology to analyse huge volumes of data had soared in the last decade. Typing ``machine learning astronomy'' into the ADS search window brings up about 20 articles for the first five months of 2016 alone, on subjects as diverse as photometric and gamma-ray burst redshift estimation, detection of radio transients, glitches in gravitational wave detection and exoplanet science \citep[e.g.][]{hoyle16,devineetal16,ukwattaetal16,zevinspy16,ford16}.

The {\it Gaia} mission\footnote{\texttt{http://www.cosmos.esa.int/web/Gaia/home}} \citep{DPACP-8} is one of the current large surveys. A cornerstone mission of the European Space Agency, it is producing repeated astrometric and photometric observations about more than 1 billion celestial objects, and opening an unprecedented insight into the  structure and history of the Milky Way. The time series of on average 70 observations per object will also allow the detection and analysis of variable celestial phenomena \citep{eyeretal15}. We expect about a 100 million objects showing stochastic, temporally localized or periodic variability, among the latter several millions of eclipsing binaries \citep{eyeretal00,holletal14}. The analysis of eclipsing binaries yields important constraints on stellar parameters at different stages of evolution, when combined with precise parallaxes. {\it Gaia}, with an aimed end-of-mission parallax accuracy of about 5-16 $\mu$arcseconds below $G=12$ mag, 26 $\mu$arcseconds at $G=15$ mag and 600 $\mu$arcseconds at $G=20$ mag\footnote{ for B1V stars, according to the In-Orbit Commissioning Review (July 2014); \\ \texttt{http://www.cosmos.esa.int/web/gaia/science-performance}}, and with the millions of binaries observed, will offer an unprecedented opportunity to study the galactic distributions and populations over an as yet unimaginably wide variety of objects.

However, detection completeness and the attainable accuracy of any derived information is highly influenced by the {\it Gaia} time sampling. This is quite sparse: the above mentioned 70 observations will be made over the course of five years, and will vary from about 40 to 240 depending on sky position. 
As a consequence, there will be a lower detection probability for systems with very narrow eclipses, depending on their position. Moreover, many details of the light curves may be missed, such as total eclipses, the exact geometry of the inter-eclipse phases or the eclipses, the start and end of eclipses, signs of third bodies in the systems or intrinsic variability of a component. In addition, the automated statistical learning methods required by datasets of millions of objects are never as good as procedures manually tailored to each object and supervised one-by-one. It is therefore important to assess what useful information can be extracted using automated methods from the  {\it Gaia}-observed light curves, what the performances of these methods are in the procuration of this information, and whether the quality of this information is sufficient to be used as aid for initial model selection in the planned detailed physical analysis of {\it Gaia} eclipsing binaries.

We study here the potential role of a few machine learning methods in this task, in particular, functional principal component analysis (FPCA), linear discriminant analysis (LDA), random forest, and self-organizing maps (SOM). To assess the performance of the methods, we use three data sets: (1) the Catalog and Atlas of Eclipsing Binaries\footnote{\texttt{http://caleb.eastern.edu}, maintained by D. H. Bradstreet, Eastern University, US} ({\it CALEB}) data, which contains 306 physically modelled eclipsing binaries extracted from the literature, and which gives not only classification into system morphology classes from detached to overcontact, but also the results of a detailed modelling based on literature; (2) 521 eclipsing binaries from the {\it Hipparcos} mission\footnote{\texttt{http://www.cosmos.esa.int/web/Hipparcos; http://casu.ast.cam.ac.uk/casuadc}} \citep{esahipparcos}, which were classified manually (using literature where available) into light curve morphology classes Algol-type (EA), $\beta$ Lyrae-type (EB) and W UMa-type (EW);  and (3) light curves of the eclipsing binaries detected by {\it Kepler} \citep{prsaetal11, slawsonetal11} resampled according to the {\it Gaia} observing times (Kochoska et al., under revision for A\&A). The first two serve partly as training sets, and partly to anchor the results of new methodology to current knowledge through the use of objects that were already analysed, checked and discussed. Hipparcos is used to assess the performance of classification into the light curve morphology classes EA, EB and EW, while CALEB, into the system morphology classes detached, semi-detached and overcontact. The third dataset consists of the most complete sample of eclipsing binaries in a small region of the sky to date. It is also endowed with the results from visual inspection and from the application of artificial intelligence for their physical parameters and system morphology class. Since for our goals, we subsampled the light curves as {\it Gaia} will observe them, this provides checks and comparisons of the performance and the biases of the methodology when applied to {\it Gaia} data.

In this paper, we suppose that eclipsing binaries have been identified by supervised classification and their period has been estimated \citep[][]{dubathetal11, rimoldinietal12, DPACP-15}. First, we implement a three-step pre-processing of the folded light curves, which produces interpolated and standardized light curves that are sufficiently robust against effects of time-sampling variations. Second,  we consider the sub-classification of eclipsing binaries by random forest and LDA into the two class systems mentioned above (the first comprising EA, EB, and EW and the second comprising detached, semidetached, and overcontact). Third, we investigate data-driven ways to explore natural low-dimensional surfaces traced by eclipsing binary light curves in the high-dimensional space of all continuous curves in the phase space. We inspect the clustering of the known objects on these surfaces. Moreover, we investigate systematic biases when the sample of {\it Gaia}-observed {\it Kepler} binaries is projected onto them (Kochoska et al., under revision for A\&A). A second paper (hereafter, paper II) will focus on the possibility of applying statistical methodology to learn an approximate link between our light curve decomposition and the physical parameters of the systems, using the objects from {\it CALEB}. 

We present the data in detail in Sec. \ref{sec:data}. Section \ref{sec:methods} gives a brief summary about the terminology and statistical methodology used, time series pre-processing, and the supervised and the unsupervised classification methods. We also give an account of how uncertainty of class estimates due to specific training set selection is derived, and how the model complexity was chosen. We discuss the results in Sec. \ref{sec:results}. We devote detailed analysis to the light curve fitting and decomposition by PCA (Sec. \ref{subsec:fpca}), to supervised  classification (Sec. \ref{subsec:class}), and to data-driven dimension reduction methods, namely, the LDA (Sec. \ref{subsec:ld1}) and SOM (Sec. \ref{subsec:som}). Finally, in Sec. \ref{sec:conclusions} we summarize the results and draw conclusions.


\section{Data} \label{sec:data}

\subsection{\it Hipparcos}  \label{subsec:hip}

{\it Hipparcos} \citep{perrymanetal97} was an astrometric and photometric satellite mission in operation between 1989 and 1993, measuring more than 100,000 bright objects on average about 110 times during the 3.5 years of the mission. The final catalogue contains high-precision photometry for 118,204 objects, of which variability analysis detected 11,597 periodic, non-periodic and micro-variables \citep[][Vol. 11]{esahipparcos}. Detailed studies of the periodic variable stars  \citep{eyer98} identified 917 eclipsing binary systems. Using visual inspection, Fourier analysis, and literature information where this was available, these were further classified manually into EA, EB and EW types according to the definition given in the General Catalog of Variable Stars\footnote{\texttt{http://www.sai.msu.su/gcvs/gcvs/iii/vartype.htm}; the site became unavailable in fall 2016.} \citep{eyer98}. A total of 521 objects were selected through the visual inspection of light curves and included well-sampled, high signal-to-noise objects, as well as those with reduced quality (fewer measurements in eclipse, lower S/N ratios, scatter of residuals beyond the level of uncertainties, small gaps in the folded light curves), provided that their eclipsing binary classification was clear from the data. In order to include the most recent information about these systems, we used in this study the subclasses from the Variable Star Index \citep[][maintained by the American Association of Variable Star Observers]{watsonetal11}, resulting in 325 EA, 129 EB and 67 EW systems. Classification into light curve morphology classes is analysed in this paper using this data set.

\subsection{{\it CALEB}}  \label{subsec:CALEB}

Classification into system morphology classes needs a set of known, thoroughly analysed systems with well-controlled physical model fits that support the classification, but this is very hard to obtain. Analyses of eclipsing binaries that satisfy the quality requirements for a reliable training set are mostly dispersed in the literature, and need time-consuming compilation and various checks. 

The Catalog and Atlas of Eclipsing Binaries ({\it CALEB}) provides such a compilation. It offers multiband photometric and radial velocity data as well as basic astrophysical parameters and literature references for 306 eclipsing binary systems. The website joins also the results of a physical modelling of the systems\footnote{The modelling was performed by the commercial software \\ BinaryMaker; \texttt{http://www.binarymaker.com}}, whose parameters were based on fits published in the literature, and the resulting system morphology classes. The parameters include estimates of orbital parameters of the systems, mass and temperature ratios, radii of the stars in different directions and the position of the L1 point, and indicators of the presence of a third body in the system (third light) and spots,  as well as other parameters necessary in the fits. We exploited this information in this paper, together with the light curves that were used to obtain the fitted parameters.

However, since the majority of the systems were chosen from those discussed and modelled in the literature, neither the passband, nor the data type (normal points, or differential photometry using various reduction methods), nor the datafile format is uniform over the collection. Photometric errors are given for only a few light curves. We compiled data from one selected photometric band of each of 294 systems from {\it CALEB}. The order of preference for passband selection was (1) Johnson--Cousins $V$ band if available; (2) Stromgren $y$ if the observations were made using the Stromgren system; (3) other Johnson--Cousins bands ($B$, $I$); (4) other passbands. 

The system morphology classes given in {\it CALEB} are detached, semi-detached, near-contact, contact, overcontact and double contact\footnote{For the definitions, see \\ \texttt{http://caleb.eastern.edu/binary\_type\_definitions.php}}. The double contact class contains only two objects. Though these systems are physically contact systems, they exhibit detached- or semidetached-like light curves due to their fast non-synchronized rotation. This peculiar association between system morphology and observable features should be learned by the machine learning methods, which is impossible with only two objects. Consequently, we omitted these objects from the used data and the class from the analysis. The semi-detached, near-contact and contact classes were merged. Beside the fact that the contact class is very small (seven objects), these classes can have similar light curve shapes, which hampers their classification, and supports the pooling of these classes. Moreover, these systems have some important common features in their configuration: they all have at least one component that attained its inner Lagrangian surface, but neither has yet overfilled it, which imprints some similarity on the light curves. We will denote detached systems with D, the pooled semidetached--near-contact--contact class with SDC, and the overcontact binaries with OC. We analyse  classification into these classes using the CALEB data.

\subsection{Overlap of {\it Hipparcos} and {\it CALEB}} \label{subsec:overlap}

\begin{figure}[!h]
\centering
\includegraphics[width=0.4\hsize]{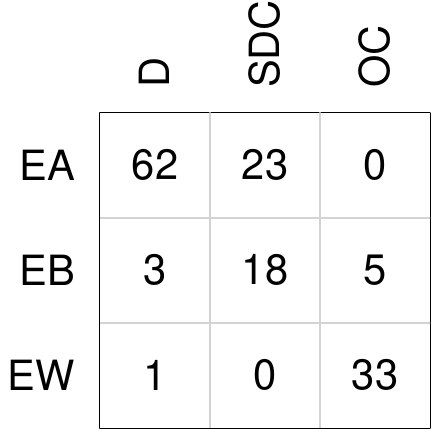}
\caption{Light curve morphology classes versus system morphology classes for the intersection of the {\it CALEB} and {\it Hipparcos} dataset.
}
\label{fig:table_eaebew_dsdoc}
\end{figure}

Our {\it CALEB} compilation contains 145 objects that have observations and light curve morphology class in our  {\it Hipparcos} selection too. Figure \ref{fig:table_eaebew_dsdoc} shows the repartition of objects according to their double classification, demonstrating  the lack of one-to-one mapping between the light curve shape class and system morphology class. In addition to the subjectivity of the visual EA/EB/EW classification, finer details of the light curves that distinguish between classes can be masked by observational noise or other effects such as pulsational variability of one of the components or circumstellar matter around the stars. One example of this is the inter-eclipse pattern of the light curves, which plays a role in the distinction between the EA and EB classes. Depending on its exact geometry, the inter-eclipse pattern of the light curves can also be a suggestion of various physical effects such as tidal distortions, reflections, or that one of the components filled its Roche-lobe. A semi-detached system falling in the latter case is AB Cas, where the $\delta$ Scuti-type pulsations of one of the stars complicate the detection of the precise shape of the light curve \citep{soyduganetal03}. Case studies of stars in the non-diagonal elements of Figure \ref{fig:table_eaebew_dsdoc} can be found, among many others, in \citet{mardirossianetal80, popper76, linneletal98, torresetal10}.

A majority of the intersection between CALEB and Hipparcos, 120 objects are used as a training set for PCA. This set contains 60 EA-type, 26 EB-type and 34 EW-type systems according to light curve morphology classification, and of 49 D, 33 SDC (16 semi-detached, 15 near-contact, and 2 contact systems grouped together), and 38 OC systems according to the system morphology classification. A total  of 25 EA-type systems were omitted in order to avoid the overwhelming dominance of this class, and thus to achieve a modelling better adapted to all classes. Once thus determined, the set of basis functions remained fixed for all further studies.

All the above datasets are small, and while sufficient for the analysis of up to a few thousand binaries, it will  be far too small to train a model for terabyte-scale survey data, where the far tails of the statistical distribution of the classes must be known for a good classification of objects that were rare or non-sampled in smaller surveys. Larger databases of light curves that are also publicly available may also be found in the literature. However, these are usually less thoroughly analysed, and carry less reliable additional information about the physics of the binaries than {\it CALEB}. Since the goal of our study is to assess the capability of different machine-learning tools to extract reliable information from light curve data, we need a control sample against which we can check machine-learning results. So we opted for using small datasets with ample information rather than large, but less well known datasets. This said, it should not be forgotten that mistakes are possible in even the best-analysed data sets. Our data may also contain misclassifications and erroneously fitted parameters. The results presented here must be considered with these caveats.

\subsection{{\it Kepler} eclipsing binaries resampled with {\it Gaia} time sampling} \label{subsec:kic}

The {\it Kepler} mission\footnote{\texttt{http://kepler.nasa.gov}} \citep{boruckietal04} observed about 160,000 sources with a $\sim$92\% duty cycle. Its 115 deg$^2$ field of view was observed from 2009 to 2013; its targets were observed in long cadence (30 min) and/or short cadence (1 min). \citet{prsaetal11} published the first catalogue of 1879 eclipsing binary systems based on Q0 and Q1 data. The eclipsing binary light curves were made publicly available\footnote{\texttt{http://keplerEBs.villanova.edu}}, along with their ephemerides, system morphology class and principal physical binary parameters estimated by a neural network. \citet{slawsonetal11} updates this catalogue with Q2 data and \citet{kirk2016} updates the catalogue for the entire mission data span.

We used this database to check the portability of our models from the mostly densely sampled {\it CALEB} data and the well-analysed, clean {\it Hipparcos} solar neighbourhood systems to a realistic population of eclipsing binaries as seen by the {\it Gaia} spacecraft. In order to achieve this, the {\it Kepler} light curves were extrapolated and resampled  according to the {\it Gaia} scanning law using AGISLab (five years of nominal mission in the {\it Kepler} field; Holl et al., in preparation, Kochoska et al., under revision for A\&A), and submitted to the time-series preprocessing and the trained classifiers. 

The dense time sampling and the photometric precision of {\it Kepler} allowed the detection of the most complete sample of eclipsing binaries of a region of the sky to date, including the detection of very narrow or shallow eclipses. Thus, the {\it Kepler} sample extends to these extreme light curve types, and it has markedly different subclass composition than {\it Hipparcos} or {\it CALEB}. Moreover, under {\it Gaia} time sampling, many of these objects with these extreme light curves have only one or two observations in the eclipses, which can hamper both light curve modelling and physical modelling. The {\it Gaia}-resampled {\it Kepler} data are used for the assessment of the biases of models trained on {\it CALEB} and {\it Hipparcos} but applied on {\it Gaia}-like time series.

\section{Statistical methodology} \label{sec:methods}

\subsection{Terminology} \label{subsec:term}

In this paper (and the subsequent paper II), we will apply some terms in a strict sense for clarity.
\begin{description}
\item[`Physical modelling':] the modelling of the binary by a code that is based on the simulation of the binary systems (configuration of the system and the surface of the stars). The best-known example is the Wilson-Devinney code \citep[][]{wilsondevinney71}. 
\item[`Light curve modelling':] a mathematical approximation to the observed light curve, such as the commonly used harmonic modelling or the combined double Gaussian--smoothing spline fit applied here.
\item[`System morphology':] the geometric configuration of the system, related to its evolutionary state, which is not directly observable. By system morphology classes, we always mean the broad types of configurations, namely, detached, semi-detached, and overcontact types. The semi-detached class includes those characterized as near-contact and contact by {\it CALEB}. 
\item[`Light curve morphology':] the shape of the observable light curves. By light curve morphology classes, that is, EA, EB and EW, we refer to the phenomenological types as defined for example in the General Catalog of Variable Stars$^4$, without direct reference to the physical type of the system. Although this classification originally intended to relate to the physical type of the system, the improvements in the physical modelling codes over decades shed light on a degeneracy between system morphologies and light curve shapes. It became clear that there is no one-to-one correspondence between the system morphology classes and the light curve morphology classes. We therefore treat them separately, with the two different meanings defined above. For further discussion, we refer the reader to the forthcoming paper of Mowlavi et al. (in preparation).
\end{description}

\subsection{Time series preprocessing} \label{subsec:tspreproc}

In our study, we use the period given in the respective databases. In practice, this means that we suppose we know the period reliably (the question of obtaining a period for eclipsing binaries will be discussed in Holl et al., in preparation). 

{\it Gaia} data and the time series from the {\it Hipparcos} and {\it CALEB} catalogues are all comprised of a few tens to a few hundred data points, though the sampling cadences are different. This implies that we need to deal with both possible gaps in phase, and the diversity of features in eclipsing binary light curves. The simultaneous presence of long, flat inter-eclipse parts and deep, narrow eclipses with high curvature at the minima within the same light curve poses a difficulty  for many curve-fitting procedures. For instance, harmonic decomposition or spline smoothing are both sensitive to it. Tuning the methods to be able to trace the eclipses or the finer details of the inter-eclipse regions would almost inevitably result in an overfitting of the noise in flat, featureless inter-eclipse regions. This is further aggravated by the sparse time sampling of {\it Gaia}, which can entail large phase gaps in the folded light curves and thus introduce uncontrolled large deviations of the harmonic fits or the splines. We therefore applied a multi-step procedure to smooth, de-noise and interpolate the folded light curves, each step aimed at dealing with different features in the light curves.

\begin{description}
\item{\textit{Double Gaussian fit.}}
We fitted the two eclipses with a combination of two Gaussian density functions (Mowlavi et al., in preparation). The Gaussian profiles model very flexibly both very narrow, sharp eclipses and broad sinusoid-like ones.
\item{\textit{Spline smoothing.}}
After the removal of most of the typical variations of the light curves by the double Gaussian fit, periodic B-splines are then able to model most of the remaining systematic variability. We found that the tuning parameter of the splines (the degrees of freedom, or equivalently, the smoothing parameter) is best to define by cross-validation within a restricted interval (the parameter \texttt{spar} in the R procedure  \texttt{smooth.spline} was set to remain within the interval $[0.35,0.9]$, and this worked well for all {\it CALEB}, {\it Hipparcos} and simulated {\it Gaia}-like data). Based on these first two steps, we shifted the primary (deeper) minimum of the light curves to phase zero and we scaled them to have amplitude one (so all light curves had scaled magnitude one and phase zero at the brightness minimum and scaled magnitude 0 at the brightness maximum). Finally, we interpolated them to obtain their values at phases $0.01,0.02,\ldots, 0.99$, forming a 99-component random vector corresponding to each light curve. 
\item{\textit{Functional principal component analysis.}}
 In the next stage, these random vectors underwent a principal component analysis. The model was trained on a selected subset, the overlap of CALEB and Hipparcos. PCA is usually applied to centred data, so we started by removing the pointwise mean (an average light curve over the used population) from the 99-component random vector. Then, on these centred random vectors (the residuals around the average light curve) principal component analysis is performed. This is equivalent to the iterative application of two alternating steps. First, the direction of maximal variance is selected in the 99-dimensional space (this is the first principal component, PC1). Then the projection to this direction is subtracted from the random vectors, reducing them thus to a 98-dimensional space. These two steps are subsequently applied iteratively on the successive subspaces, giving rise to a series of principal component vectors  \citep[PCs; see, e.g.][]{jolliffe,hastieetal}. 
 
 When applied to functions represented by random vectors \citep{ramsaysilverman1, ramsaysilverman2}, these PCs determine a series of basis functions which can be used to decompose any light curve. The order of the basis functions is such that the variance of the decomposition coefficients of the first one (the coefficient of PC1) is the largest over the training set, decreasing monotonically towards higher PC orders. Thus, the statistical variance of all light curves in the examined data set is decomposed into the variances of orthogonal projections. A few applications of PCA and FPCA in astronomy are \citet{paltaniturler03, kanburmariani04, savanovstrassmeier08, debsingh09, suvegesetal12b, zhaoetal16}.  In our case, the basis functions can be used for a perfect reconstruction of any fitted and interpolated light curve. However, since the variance of the coefficients decreases with order, it can be supposed that the dominant, systematic, most characteristic and most frequent variation modes in eclipsing binary light curves will be reflected by the first few PCs, while less frequent, tiny or non-systematic types of variation and random fluctuations due to noise or wrong fits will be modelled by higher-order terms. Effects of the typical observing cadences in the surveys (daily for {\it CALEB}, 6 hours for {\it Gaia}) or other quasi-periodicities (e.g. persistent common weather patterns, intrinsic variability of one component of the binary) are expected to appear as noise after folding the light curve with its period (or even with an alias thereof). The only case when these effects would be clearly visible when their period is commensurable with the period of the binary. However, these cases usually show specific patterns, and can be recognized at the classification preceding the selection of eclipsing binary candidates from a survey. Consequently, FPCA can serve as both a dimension reduction method for capturing the essential elements of binary variability, and as a general de-noising method. This makes it a good candidate to produce inputs for classification methods, where the leading variation modes can be expected to be informative about the system or light curve morphology class.
\end{description}

\subsection{Classification} \label{subsec:classif}

After creating the principal component basis using the overlap of {\it Hipparcos} and {\it CALEB} data sets, we projected all the smoothed-standardized light curves from {\it Hipparcos}, {\it CALEB} and the {\it Gaia}-resampled {\it Kepler} eclipsing binaries onto these basis functions, and extracted the coefficients PC1, PC2, and so on, of the basis functions. The PC coefficients fully characterize the shape of the light curves. Together with the period and the amplitude, they comprise the complete information available in the single-band photometric light curve, presumably cleaned from noise, as an appropriate input for machine learning methods.

We applied several classification techniques, both supervised and unsupervised, to perform classification into the commonly used class systems and to discover natural low-dimensional structure in the complex manifold of light curves.
\begin{description}
\item[\textit{Linear discriminant analysis}] \citep[LDA;][]{mardiakentbibby} is a basic supervised classification tool that determines linear boundaries between classes, and gives a simple class estimate of each data instance. It assumes multivariate Gaussian probability distributions in the attribute space for each class, with identical covariance matrix but centred at different positions. It finds estimates of the centres and the class covariance matrix such that the variance of the class centres is maximized, while the within-group variance in each class is minimized. Despite the apparently very restrictive assumption, LDA performs well on the task of eclipsing binary classification. Moreover, we find that the score of the systems on LD1, the first discriminant function, yields a very well-behaved characterization parameter, similar in nature to the morphology parameter of \citet{matijevicetal12}, and maps thus the light curves of the stars to a continuous one-dimensional manifold. 
\item[\textit{Random forest}] \citep{breiman01} is a very popular tree-based supervised learning method which performs very well in a large range of classification problems. The ``forest'' consists of many trees, each of which is grown using a double randomization: for each tree, we select randomly a subset of the training set, and at each node, we first randomly select a small subset of all attributes, then use the best split-point of only these to split the node and to grow the tree. Random forest offers some advantages, detailed below.
\begin{itemize}
\item This two-level randomization is an efficient way to reduce the variance of the aggregated estimate of the forest and to avoid overfitting.
\item Random forest is also applicable in high-dimensional classification problems, that is, with the number of training instances being smaller than the number of attributes. Each tree considers the classification problem in a much lower-dimensional space, and thereby avoids the curse of dimensionality, namely, data becoming exponentially sparse in spaces of increasing dimensions.
\item Growing a high number of trees can be used to extract information on uncertainty due to sampling variance: for any object, each tree in the forest casts a (possibly different) vote for one of the classes, and these votes can be aggregated to obtain relative frequencies of class labels. This can be considered as a posterior probability distribution of class labels, summarizing our uncertainty about the label due to the sampling variance of the training set.
\item For each tree, there is a subset of the training data that was not used in the construction of the tree (the out-of-bag sample), and this offers both a built-in control of the performance on a test set, and a way to assess the importance of each of the attributes.
\item It can be also adapted to a strongly imbalanced class composition: specific loss functions can be prescribed to favour classification into small classes, and thus help the extraction of an interesting but rare class from a population dominated by other large classes. We will use this feature in paper II, to identify systems where inclusion of spots or third light could help physical modelling, since such systems amount to only about 1/9 of the {\it CALEB} data set.
\end{itemize}
Random forest has been successfully applied for many problems in astronomy \citep[see, e.g.][]{dubathetal11, richardsetal11, rimoldinietal12, kimetal14, goldsteinetal15}. For our classification tasks, we grew all the trees to maximal extent (pure leaves or containing only one data point), we randomly sampled between three and six attributes to test at each node for the best split (depending on the model complexity), and grew 2000 trees in each fit (the error rates became always stable between 500 and 1000 trees).
\item[\textit{Self-organizing maps}] \citep[SOM;][]{kohonen90, kohonenetal00} are an unsupervised dimension reduction method, mapping data living in high-dimensional attribute spaces to a low-dimensional, possibly nonlinear, embedded subspace. The procedure is  initialized by taking a (most commonly rectangular or hexagonal) grid of  ``prototypes'' in a  low-dimensional subspace, and then, considering all data points one after the other, gradually move the neighbouring prototypes towards the data points by some fraction of their distance from it. This fraction is decreased during this procedure, until the position of the prototypes is stabilized. The result usually depends on the selected geometry of the grid, the definition of ``neighbourhood'' and the learning rate. Nevertheless, as the procedure is completely data-driven, and there is no preliminary knowledge injected in the form of class labels, it can reveal unknown intrinsic low-dimensional structure and non-trivial connections between regions of a high-dimensional space. An example of astronomical application of the SOM is in \citet{carrascokindbrunner14b}, where it serves as a tool for the estimation of photometric redshifts. 

In this paper, we used a two-dimensional $12 \times 8$ rectangular grid, and we defined the neighbourhoods as fixed-radius balls. The parameters of the learning rate were optimized through the minimization of the error, that is, the difference between the true attribute values of the objects and the attribute values of the closest gridpoint (their ``cell''). No attribute other than the PCs was used, so the SOM was solely based on light curve shape. The SOM model was trained with all {\it CALEB} data, then the {\it Gaia}-sampled {\it Kepler} eclipsing binaries were projected on the found surface.
\end{description}

\subsection{Checking the classifiers} \label{subsec:modelcheck}

The assessment of the quality of the statistical models and diagnostics for possible problems are just as important stages of data analysis as model building itself, since this is what provides information about the precision and reliability of our model. We applied randomization to confirm the usefulness of the classification models and to obtain inference in the form of posterior distribution functions for point estimates. 

Firstly, in order to obtain posterior probabilities for the class labels in the supervised classification, we randomize the training sets: we randomly select 150 objects from our data sets 500 times, and construct models on each of the 500 sets. LDA gives only a point estimate for the class label. We used the empirical distribution function of the 500 estimates to form an idea about the posterior distribution function of the class labels. However, though this procedure accounts for the variations due to the training set selection, it does not propagate the uncertainty of the attributes to the final estimates.

Secondly, in order to assess whether the fitted model brings a real improvement due to the use of information inherent in the used attributes, we resampled the values of the class labels in the training set with replacement. This is a nonparametric bootstrap, a viable solution in the case when parametric simulations cannot be used because the true distribution of the parameter of interest in the population is unknown. We thereby break any true correlation or dependency between the class labels and the attributes. We did this bootstrap for each of the above described training set selections, and repeated model fitting (both with LDA and with Random forest) on these doubly randomized training sets. Comparing the results obtained with the randomized and the non-randomized data sets informs us  whether the result of the classification is just a random outcome, not significantly better than a random label allocation, or the attributes indeed contain information about the class labels.

\subsection{Attribute selection} \label{subsec:varselect}

In principle, we could use all 99 principal component coefficients, complemented with the amplitude and period, in our classification procedures. However, such relatively high number of features needs to be treated particularly. Random forest, as we mentioned above, can in theory give valid results in high-dimensional setups. However, if (as is usually the case) many of the used attributes are irrelevant, the performance would be improved by dropping these features, since they would not mislead classification anymore by forcing random decisions in the trees. To select the meaningful attributes, we used the method's built-in attribute ranking capability, which consists of randomizing one by one each of the attributes over the out-of-bag sample for each tree and computing the average accuracy loss due to each. The attributes that cause the highest loss are retained in the final models.

The curse of dimensionality affects the performance of LDA as well, as for all density-based clustering method. Thus, we performed LDA in the subspace of the 25 leading PCs at most. The selection of the optimal model complexity was done based on the classification scores for models constructed in the successive subspaces of the first two, three, four, five, eight, eleven, fifteen and twenty-five PCs. 

All the analysis presented in this paper was performed in the statistical environment R \citep{R}, using in particular its packages \texttt{randomForest}, \texttt{MASS} and  \texttt{som}. The computational time to train one Random Forest model with 5000 trees was below a second, for one SOM model, of the order of a minute on a 2015 2.8GHz MacBook Pro laptop. No parallel computing was used.

\section{Results and discussion} \label{sec:results}

\subsection{Pre-processing and FPCA} \label{subsec:fpca}

\begin{figure*}[!h]
\centering
\includegraphics[width=0.96\hsize]{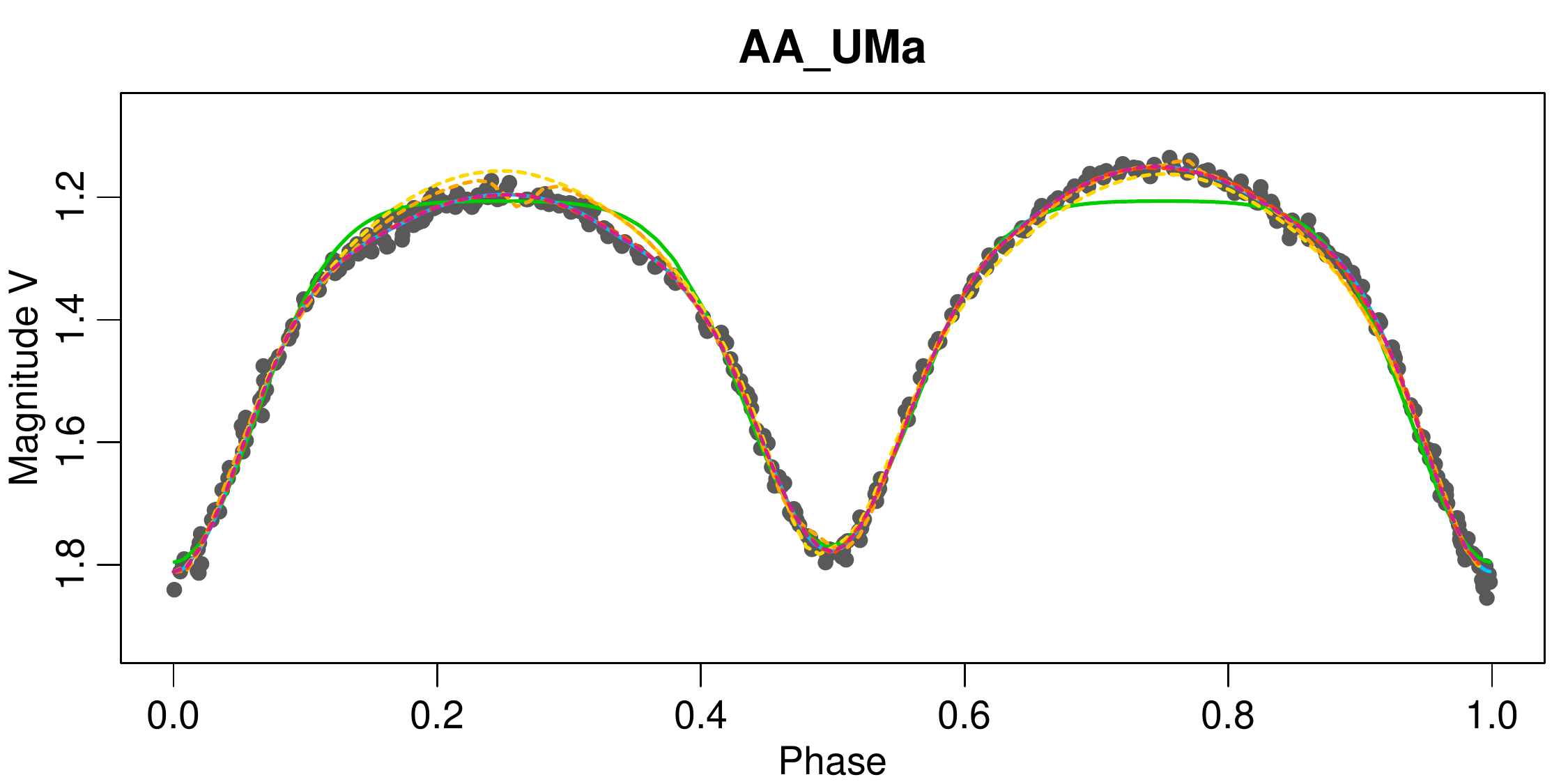}
\includegraphics[width=0.96\hsize]{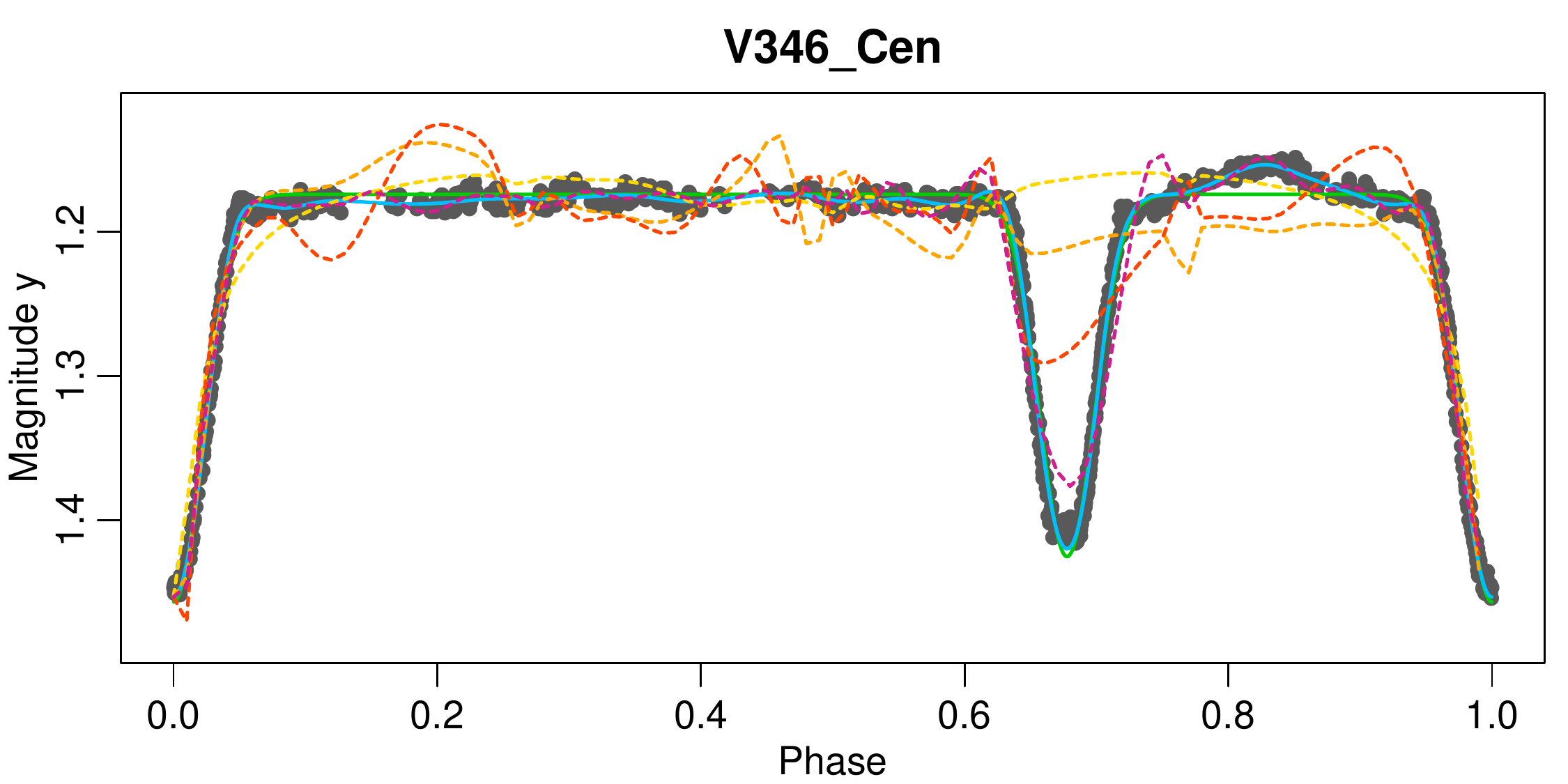}
\caption{Two examples, AA UMa (period: 0.4681258 days) and V346 Cen (period: 6.3219350 days) for the successive time series preprocessing steps (the black dots indicate the data taken from CALEB). The magnitudes for AA UMa are $V$ magnitudes from \citet{wanglu90}, for V346 Cen, Str\"omgren $y$ from  \citet{gimenezetal86}. Green solid line: the double Gaussian fit, blue solid line: double Gaussian + smooth spline fit, dashed lines: principal component reconstructions up to different orders. Gold: PC1-PC4 (95\% of the total variation), orange: PC1-PC11 (99\%), red: PC1-PC21 (99.9\%), violetred: PC1-PC36 (99.999\%). The light curves  reconstructed from the PCA decomposition are scaled back from between [0,1] to the original scale of the data.}
\label{fig:successive_lc_fits}
\end{figure*}

\begin{figure}[!h]
\centering
\includegraphics[width=\hsize]{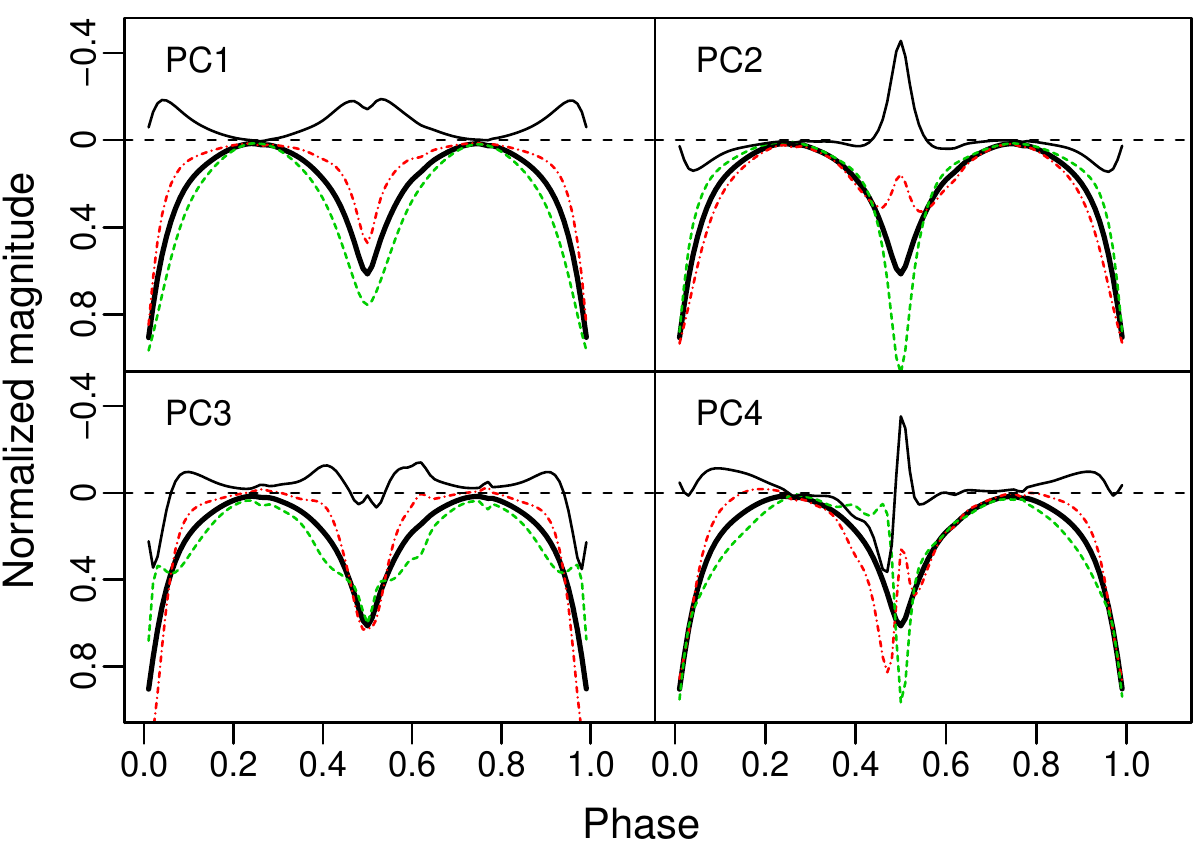}
\caption{The first four basis functions (PC1, ..., PC4) of the PC decomposition, represented with the thin black line in the upper part of each panel. The lower part shows the modification of the mean light curve when the principal component function in the upper panel is added (red dashed line) or subtracted (green dashed line).}
\label{fig:FPCAbasis}
\end{figure}

The advantages and drawbacks of the multi-stage pre-processing are illustrated in Figure \ref{fig:successive_lc_fits}. The double Gaussian fit serves as a primary modelling of the light curves. Although it does an excellent job of roughly modelling the eclipses and avoiding some of the pitfalls of Fourier analysis of sparsely sampled time series, its failure to capture the fine geometry of the light curves is noticeable in both panels, for example the asymmetric brightness of the two inter-eclipses for AA UMa. The added smooth spline fit of the residuals (almost completely obscured by the lines of PC reconstructions for AA UMa) corrects this. The upper panel illustrates the capacity of the splines to account for the inter-eclipse geometry and details of the eclipses. The lower panel shows one of the most difficult cases: despite the improvement in the inter-eclipse between phases 0.7 and 0.95, the smooth spline fit does not trace very well either the breakpoints of the likely total secondary eclipse or the endpoint of the secondary eclipse. This is the manifestation of the difficulties of any smooth modelling to trace changes where the change to be modelled is itself not smooth (its first derivative is discontinuous): changing the smoothing parameter to enable the splines to follow the sharp angles would result in overfitting of the scatter between phases 0.1 and 0.6. However, according to our further studies this loss is not crucial for the recognition of the class or for the SOM position.

For the construction of the principal component model, the vectors of the shifted, scaled, interpolated magnitudes at phases $0.01, 0.02, \ldots, 0.99$ of the 120 objects in the overlap of {\it Hipparcos} and {\it CALEB} are used. First a mean light curve is computed by taking the average of the magnitudes at each phase. Then the residual light curves, after removal of this pointwise mean, undergo the PCA. The four most important basis functions of the FPCA decomposition, together with the mean light curve, are shown in Figure \ref{fig:FPCAbasis}.

The function PC1 captures (to a first order) the variation from detached system geometry or EA-type light curve (narrow, sharp eclipses of different depths with flattish inter-eclipse regions) to overcontact system geometry or EW-like light curve shape (approximately sinusoid) with very similar eclipse depths. The secondary eclipse is triangular and shallow for large positive coefficients of PC1, and deep and rounded for large negative coefficients (both are indicated in Fig.  \ref{fig:FPCAbasis}).  The variation proportional to this basis function represents 81\% of the total variance of all the light curves. In other words, this is the function which has the coefficients varying in the widest range on the training set, and therefore can be regarded as the dominant mode of light curve variation among eclipsing binaries. We note that the fraction of variance that is modelled by the first PC component for individual light curves is not necessarily equal to this 81\%; this is a population average. As can be seen in Figure \ref{fig:successive_lc_fits}, this fraction may be even higher for AA UMa, while it must be certainly lower for V346 Cen.

The function PC2 fine-tunes the depth of the secondary minimum, and at the same time, slightly corrects for the shape of the primary minimum. The dominant element of the variation here is the variation of the secondary depth, corresponding to variations in the mass ratio with respect to the dominant one indicated by PC1. It accounts for another 8\% of the total variance.

The PC function corresponding to the third largest variability component, PC3 captures the geometry of the transition between eclipses and inter-eclipses. It leaves the depth of the secondary minimum intact (though it strongly overfits the primary minimum), but modifies the curvature or the position of the breakpoint in the slopes. The fraction of systematic variability picked up by this component is nearly 5\%.

The fourth PC function, PC4 is the first related to asymmetry of the light curves with respect to phase 0.5. It modifies the slopes of the minima (those of the primary more markedly), and at the same time, shifts slightly the position of the secondary minimum. It can therefore play a role in the modelling of eccentric systems. However, while the shape of the primary eclipse may be indeed a fairly important, frequently appearing variation in the population of binaries, we have no reason to suppose that the phase of the secondary minimum, which depends mainly on our line of sight on a given eccentric system, is associated with a particular shape of the primary minimum. The appearance of this component is therefore probably due to the training set composition, a random overrepresentation of objects where this (haphazard) association of characters is present. Its contribution to the total variance is barely more than 1\%.

Light curve components proportional to these four PCs account for 95\% of the total variance of the complete light curve population around the mean light curve. Higher-order PC terms (above PC4) contribute less than 1\% each to the total variance, and represent in majority rather a mathematical decomposition of the random diversity of the light curve shapes in the training set than further interpretable elements of the general eclipsing binary light curve varieties (as above stated, already PC4 is partially driven by the random composition of the training set). This diversity is in part due to imperfections of the fits which are random, and in part, to various phenomena that have a wide range of effects on the light curve shape. Examples of the latter are binaries with nonzero eccentricity, third light and spots. Nonzero eccentricity can be seen in the light curve as the displacement of the secondary minimum away from phase 0.5, but as the phase of the secondary can vary from near-zero to near-one, and can appear in light curves with very diverse secondary depths, FPCA cannot find a small set of basis functions to summarize its effect. The situation with third light and spots will be discussed in paper II.

The quality of the approximation using different numbers of PC basis functions can be seen in Figure \ref{fig:successive_lc_fits} through the lines from yellow to red. For the case of AA UMa (top panel), the first approximation plotted (using the first four PCs, accounting for 95\% of the total variance of the training set) cannot yet adequately model the different inter-eclipse brightness maxima. Using further terms, the quality of approximation improves, until the last one (up to PC36, plotted in Figure \ref{fig:successive_lc_fits}) becomes almost indistinguishable from the double Gaussian plus smooth spline fit. The bottom panel illustrates the difficult case of an eccentric system: the position of the secondary minimum and the sharp breakpoints in the light curve require a very high number of PCs for even a rough approximation.

\begin{figure}[!h]
\centering
\includegraphics[width=\hsize]{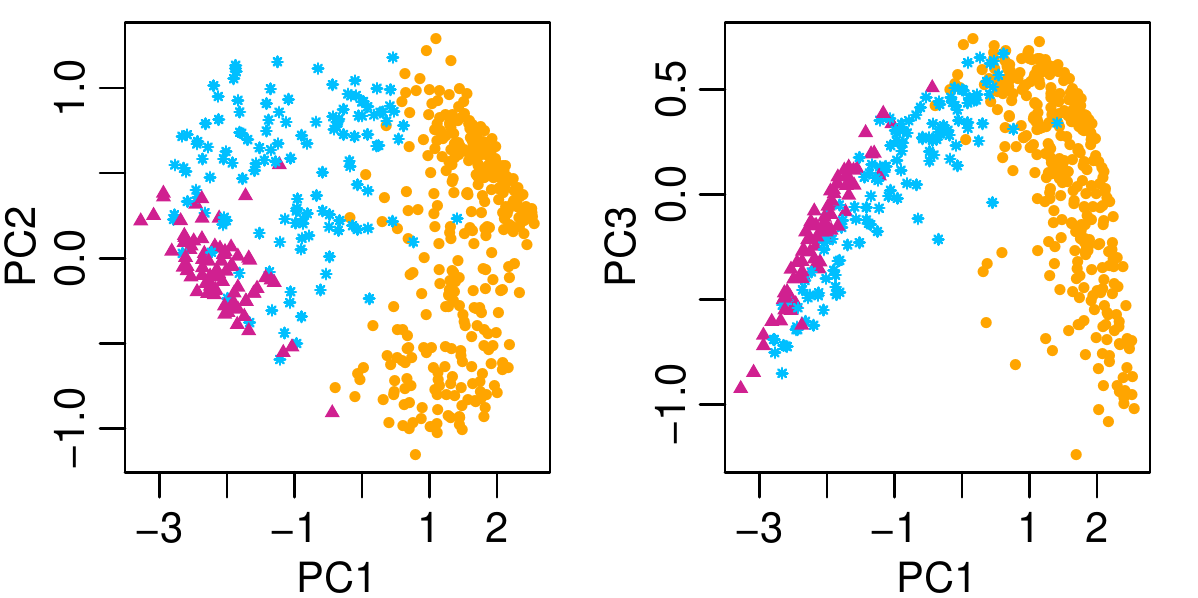}
\caption{The distribution of the EA/EB/EW classes of the {\it Hipparcos} data in the most important PC projection planes. Orange dots: EA, blue stars: EB, violetred triangles: EW.}
\label{fig:FPCscores_EAEBEW}
\end{figure}
\begin{figure}[!h]
\centering
\includegraphics[width=\hsize]{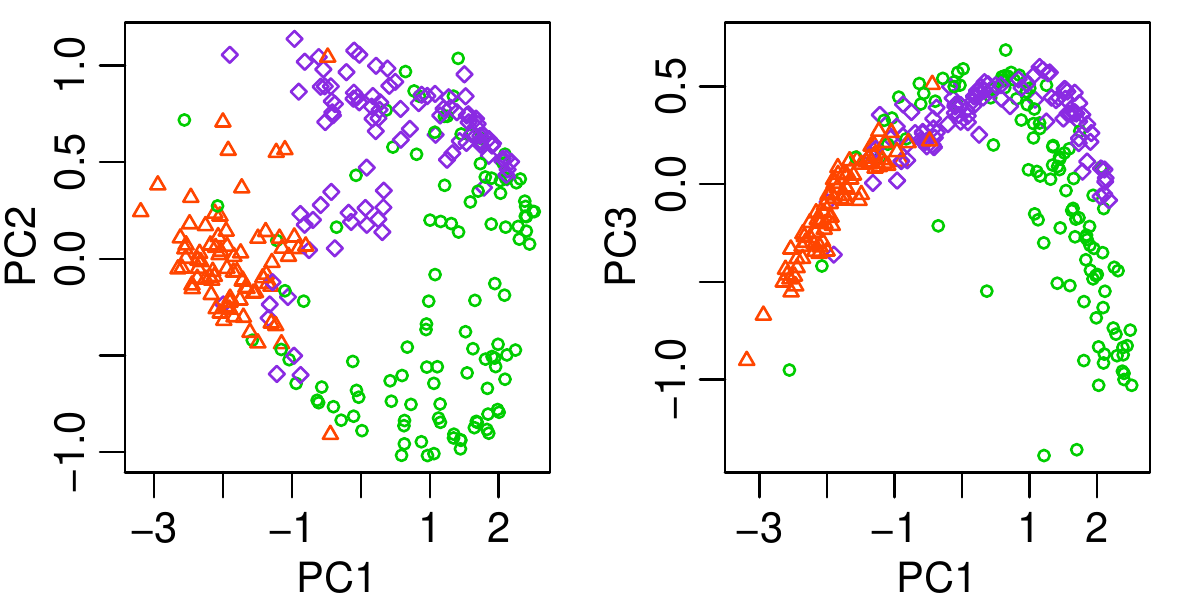}
\caption{The distribution of the system morphology classes of the {\it CALEB} database in the most important PC projection planes. Green circles: detached, violet diamonds: semi-detached (including near-contact and contact, cf. Section \ref{subsec:CALEB}), red triangles: overcontact systems.}
\label{fig:FPCscores_DSDOC}
\end{figure}

After training the principal component model, any light curve preprocessed in the same way (fitted by a double Gaussian and  spline-smoothed, shifted-scaled-interpolated) can be decomposed according to the principal component basis. The coefficients obtained on the first three PC basis elements are shown in Figures \ref{fig:FPCscores_EAEBEW} and \ref{fig:FPCscores_DSDOC}, highlighting the distribution of the light curve morphology and system morphology classes using {\it Hipparcos} and {\it CALEB} systems, respectively. The relatively tight, quite well-separated shapes of the point clouds in Fig. \ref{fig:FPCscores_EAEBEW} confirm that the leading terms of the PC decomposition indeed give an appropriate concise summary of the dominant variations in light curve shapes. 
Figure \ref{fig:FPCscores_DSDOC} reflects the degeneracy of possible system morphologies behind the light curve shapes, although some segregation can still be observed. The transition from detached through semi-detached, near-contact and contact to overcontact systems appears to be continuous, with overlaps between classes in large portions of the planes. In particular, semidetached, near-contact, and contact systems occur in almost all parts of their region mixed with other classes.  A clean separation into system morphology classes thus seems unlikely using only information from the photometric light curve.

\subsection{Supervised classification} \label{subsec:class}

\begin{figure*}[!h]
\centering
\includegraphics[width=\hsize]{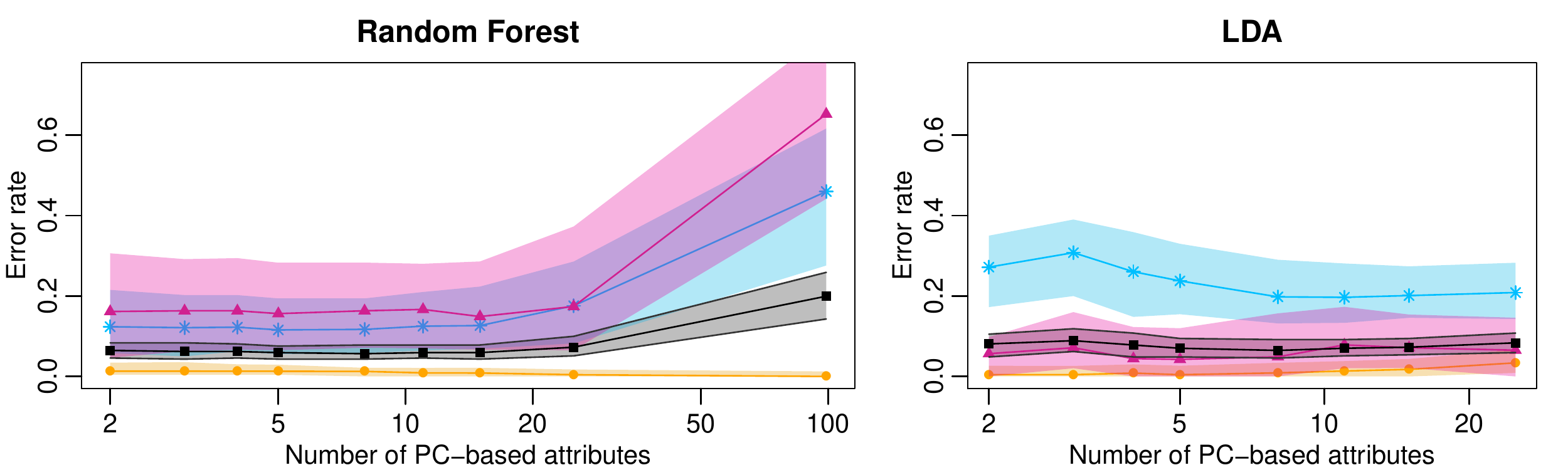}
\caption{Error rates of the ensemble members of Random forest (left) and LDA (right) for the classification into classes EA/EB/EW using \textit{Hipparcos}, as a function of the number of light curve shape attributes. The ensemble median global error rate is plotted in black (with superimposed black squares), and the range of its (0.1, 0.9) quantiles in grey, highlighted by a thin black line. The same quantities on the three classes are shown in orange (EA), blue (EB) and pink (EW). The models contain period and peak-to-peak amplitude beside the PCs, whose number is indicated in the x-axis of the plot.}
\label{fig:ensPerf_EAEBEW}
\end{figure*}

\begin{figure*}[!h]
\centering
\includegraphics[width=\hsize]{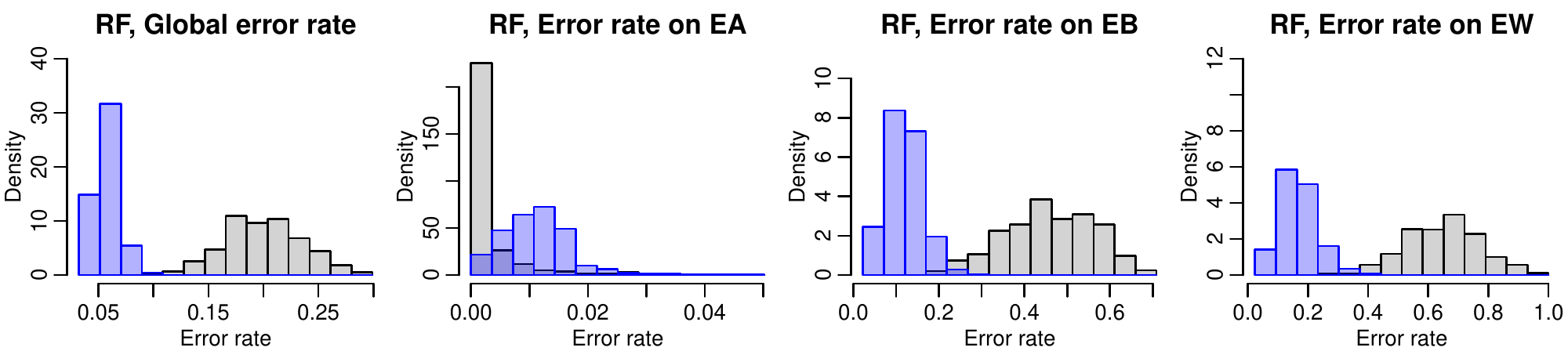}
\caption{Error rates of the Random forest classifier using the {\bf randomized} training sets in the light curve morphology classification using \textit{Hipparcos}. The error rates are measured on objects only in the test sets for each training set/test set partition. Grey histograms: period, amplitude, all PCs; blue: period, amplitude, PC1--PC8 (best performing model).}
\label{fig:ensRF_ens_EAEBEW}
\end{figure*}

\begin{figure}[!h]
\centering
\includegraphics[width=\hsize]{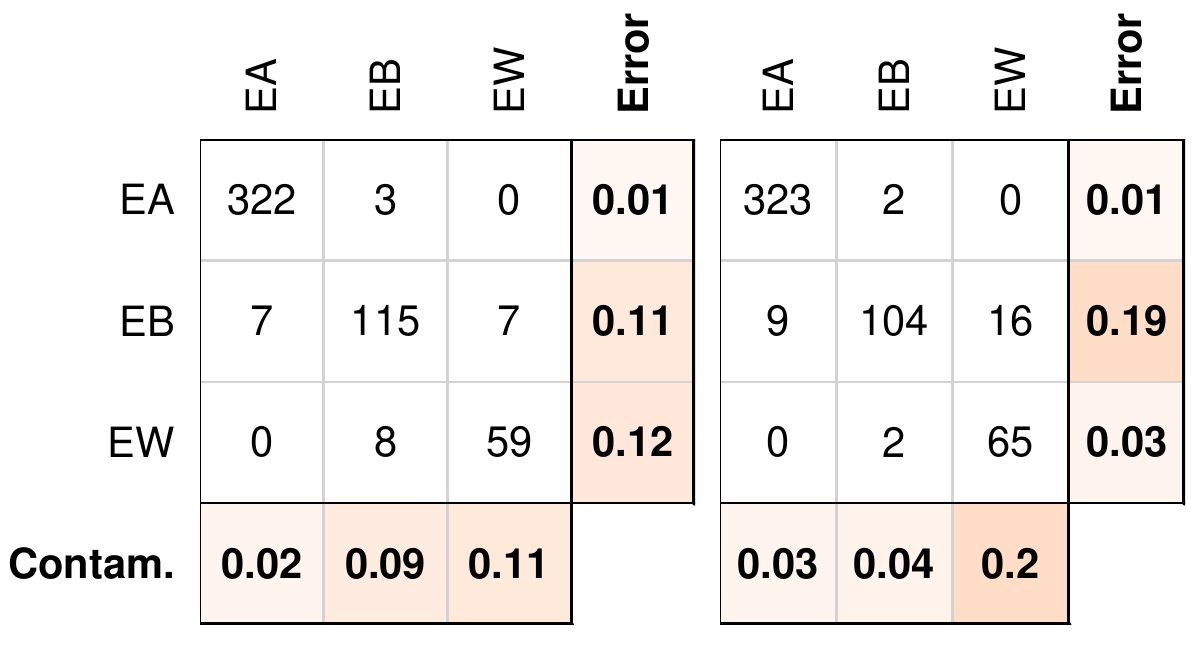}
\caption{Confusion matrices based on the aggregated results of the 500 {\bf randomized} Random Forest (left) and LDA (right) classifiers for the light curve shape classes, using the 500 random training set/test set partitions on \textit{Hipparcos}. The numbers in the white boxes are absolute numbers, those in the orange boxes are fractions (bottom row: contamination rate in the estimated class, rightmost column: error rate).}
\label{fig:confmat_EAEBEW}
\end{figure}

\begin{figure*}[!h]
\centering
\includegraphics[width=\hsize]{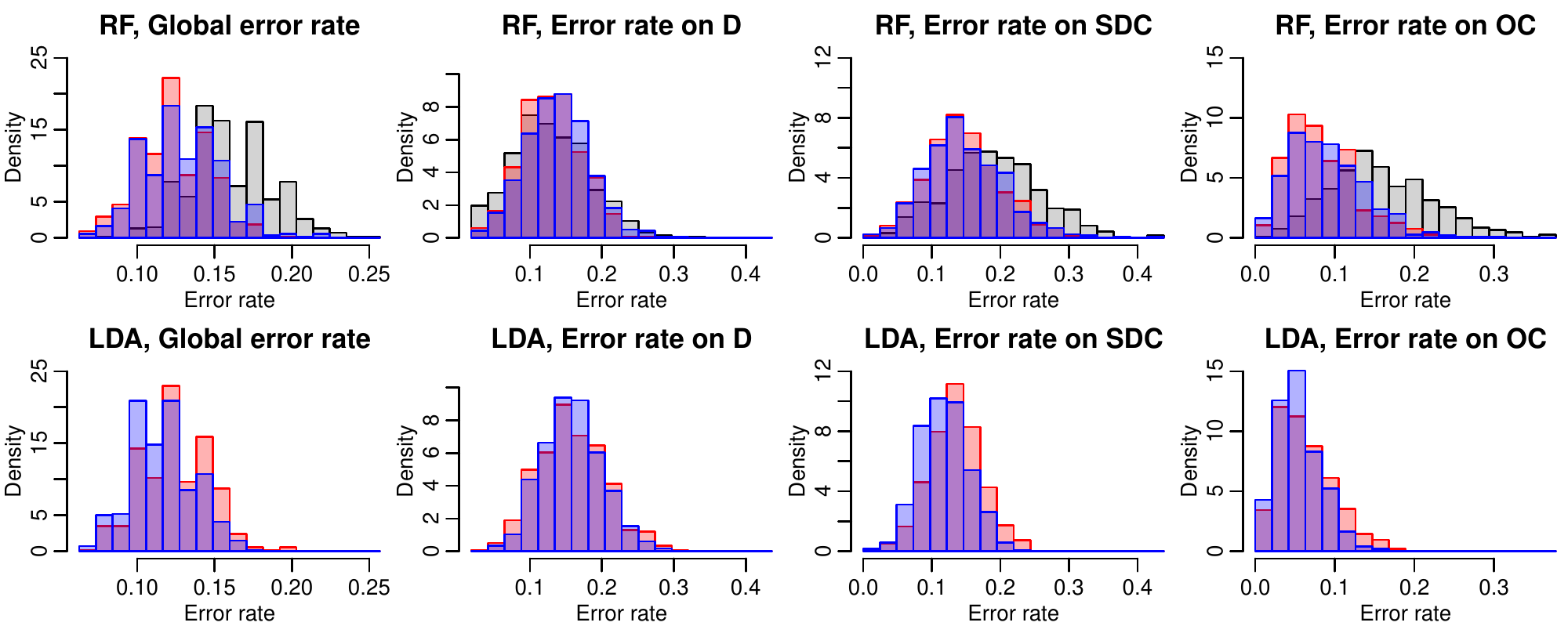}
\caption{Histogram of the error rates of the classifiers in the system morphology classification on the 500 {\bf randomized} training set/test set partitions, using \textit{CALEB}. The error rates are measured for both Random Forest (top row) and for LDA (bottom row) on objects only in the test sets for each partition. Grey: period, amplitude, all PCs (only for Random Forest), red: period, amplitude, PC1--PC11, blue: period, amplitude, PC1--PC4.}
\label{fig:ensLDA_ensRF_DSDOC}
\end{figure*}


\begin{figure}[!h]
\centering
\includegraphics[width=\hsize]{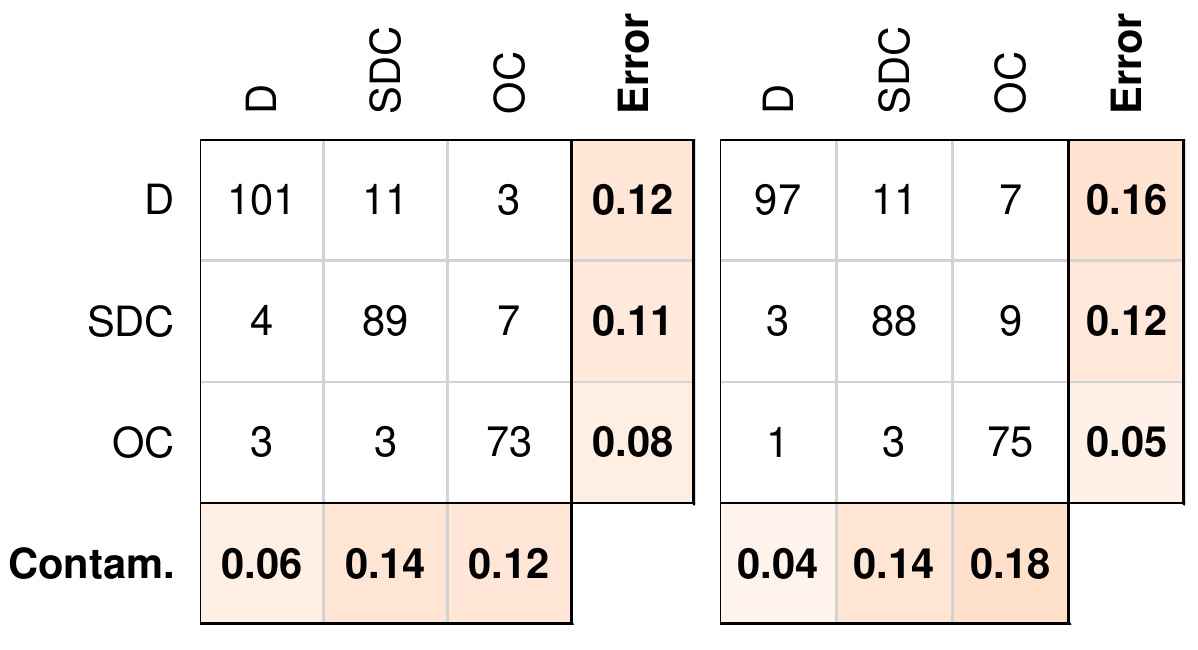}
\caption{Confusion matrices based on the aggregated results of the 500 {\bf randomized} Random Forest (left) and LDA (right) classifiers for the system morphology classes, using the 500 random training set/test set partitions on \textit{CALEB}. The numbers in the white boxes are absolute numbers, those in the orange boxes are fractions (bottom row: contamination rate in the estimated class, rightmost column: error rate).}
\label{fig:confmat_DSDOC}
\end{figure}

\subsubsection{Random forest and LDA into EA/EB/EW}\label{subsubsec:lcclass}

As a first step to find the best-performing model complexity, we ran Random forest and LDA using period, peak-to-peak amplitude and various numbers of PC coefficients from the set $\mathrm{PC1}, \ldots, \mathrm{PC99}$. For LDA, only period, amplitude and at most the first 25 PC coefficients on all 500 random training sets from {\it Hipparcos} were used; for Random forest, a full model containing period, amplitude and all PC coefficients was tried as well. The resulting global and class-wise error rates (defined as the proportion of misclassified objects among all and that of the misclassified objects of true type X, respectively) are shown in Figure~\ref{fig:ensPerf_EAEBEW} against model complexity (the number of PC attributes in the model). We concluded that Random forest models that use all terms from  $\mathrm{PC1 }$ to $\mathrm{PC5}$ are nearly equivalent regarding their overall performance to models using all lower-term PCs from $\mathrm{PC1}$ to $ \mathrm{PC15}$, with median error rates ranging from 5.6\% to 6\% over the 500 different partitions into training and validation set. There is no significant improvement if we perform a systematic backward elimination procedure using the attribute importances, either. The model selected as best uses $\mathrm{PC1}, \ldots, \mathrm{PC8}$ beside period and amplitude. This attribute set yields the overall best performance for LDA, too. 

The class-wise performances of the two classifiers are different. The class EA is extremely well distinguished by both, but EB and EW have opposite behaviour. While random forest performs uniformly around 13\% error rate on the class EB  if using fewer than 20 attributes, the error of LDA is never less than 20\% on this class, and even increases with decreasing model complexity. The class EW, on the other hand, is less well distinguished by random forest than EB. Visual inspection of Figure \ref{fig:FPCscores_EAEBEW} helps to understand this. LDA enforces a unique intra-group variance-covariance matrix for all the classes, which implies a linear separation boundaries between classes. This boundary, due to the identical intra-group variance-covariance matrices, is likely to allocate many EB systems to the EW class (those in the overlap of the classes), while it loses relatively fewer EW systems (because EWs are more concentrated around the group centre, and EBs are more dispersed). Random forest is not restricted in such a way. Its decision boundaries between the classes are not restricted to be linear, and are able to follow very intricate patterns in high-dimensional spaces. As a consequence, it is able to find a boundary through the overlap of the classes that implies better balanced loss. The numbers in the cases of the right panel of Figure \ref{fig:confmat_EAEBEW} confirm this interpretation.

Figure \ref{fig:ensRF_ens_EAEBEW} shows the behaviour of the classes, particularly that of EA in the random forest classifier in more detail. It presents the histogram of the global and class-wise error rates of two random forest models, one using all PC coefficients (in grey), and one using PC1--PC8 (in blue) in addition to period and amplitude, on the 500 distinct training set selection. While using all PC coefficients produces an excellent classification of EA objects (see the grey column close to 0 in the second panel from left), the same fails catastrophically with both EB and EW objects (the two right panels), so badly that the global error rates fall around 20\%. The explanation is that usually, the first few PC components reproduce already the most important features of the light curves of EB and EW types, and all the other, higher-order PC terms describe only noise or tiny effects unimportant for the classification. These, therefore, act as noise in the classification of these classes. However, the same high-order PCs are often relevant in the decomposition of EA systems: high values on these imply the necessity of a lot of fine corrections to obtain a light curve with sharp features or with nonzero eccentricity. The error rate of the EA systems thus slightly increases when omitting these high-order PCs from the classifier. However, the error rate on EBs and EWs improves dramatically, while it does not increase to more than 2\%  on the EA type variables.

This case of classification into light curve morphology classes is a nice example where two classifiers based on different principles perform very differently on the various classes, and therefore incites to find more sophisticated solutions. One possibility is to consider a multi-stage approach. The most promising setup consists probably of first separating  EA systems from the rest by a high-dimensional random forest classifier (since, as Figure \ref{fig:confmat_EAEBEW} shows, this leaves only up to 1\% of the true EAs misclassified, and only little contamination from EB and EW objects), then using a binary LDA or random forest classifier with fewer attributes to distinguish the two remaining classes. Such a procedure would also help a frequent problem of classifiers, namely that attributes that are very relevant to one class just induce more mistakes in the other classes. A second possibility can be the combination of the results from the different classifiers. Such solutions aim to exploit the strength of all participating methods by finding a way of combining their individual results to obtain a general improvement. Some combination procedures, with application in photometric redshift estimation, were reviewed by \citet{dahlenetal13} and \citet{carrascokindbrunner14a}.

The same models based on only the shape attributes, without the use of period and amplitude, perform in general with 1\% higher error rate. Both of these attributes have slight distributional differences over the classes. Thus, despite the fact that the EA/EB/EW classes are solely based on the shape of the light curves, classification nevertheless improves with the inclusion of these attributes unrelated to the light curve patterns.

\subsubsection{Random forest and LDA into D/SDC/OC} \label{subsubsec:evclass}

We repeated the procedure of fitting a full model, selecting relevant attributes and inspecting the class-wise performance of the classifiers with different complexity for the system morphology classification. We summarize the results in Figure \ref{fig:ensLDA_ensRF_DSDOC}. 

Although decreasing the number of attributes in a random forest classifier has a similar improving effect on the global error rates and a similarly adverse effect on the D systems compared to the SDC and OC systems as in the case of EA/EB/EW classification, the difference is not nearly so spectacular as there. The full model has somewhat lower error rates for D systems than the reduced models, and markedly higher error rates for SDC and OC. Similarly to the light curve shape classification, the performance improves on these two latter classes as well as globally with dropping the high-order PCs, but slightly decreases on D. All models that use all lower-order PCs up to a maximal order between PC5 and PC15 perform similarly: the median of the 500 global error rates is 12.5\% for all model complexities. After aggregating all the 500 results on the different training sets (just summing up the votes of all the trees from all partitions), we find that the model using PC1--PC15 performs slightly best (with a global error rate of 10.5\%). The confusion matrix of this model is shown in the left panel of Figure~\ref{fig:confmat_DSDOC}. The improvement may be due to the fact that the aggregation uses effectively the whole known set for training, while each of the 500 models used only about half of it.

The performance of LDA improves on average with dropping PC terms, until we reach a median global error rate of about 11.5\% over the 500 partitions when including the first three, four or five PCs with period and amplitude.  We can define an aggregated result on the 500 partitions for LDA too, by taking the most frequent class label as the aggregated estimator; the best performance, 11.5\% similarly to the median global error rate, is achieved by the same models using three, four or five PCs beside the period and amplitude. The model with PC1--PC4, period and amplitude is shown in the right panel of Figure~\ref{fig:confmat_DSDOC}.

\subsection{Transition to data-driven clustering: LD1} \label{subsec:ld1}

\begin{figure}[!h]
\centering
\includegraphics[width=\hsize]{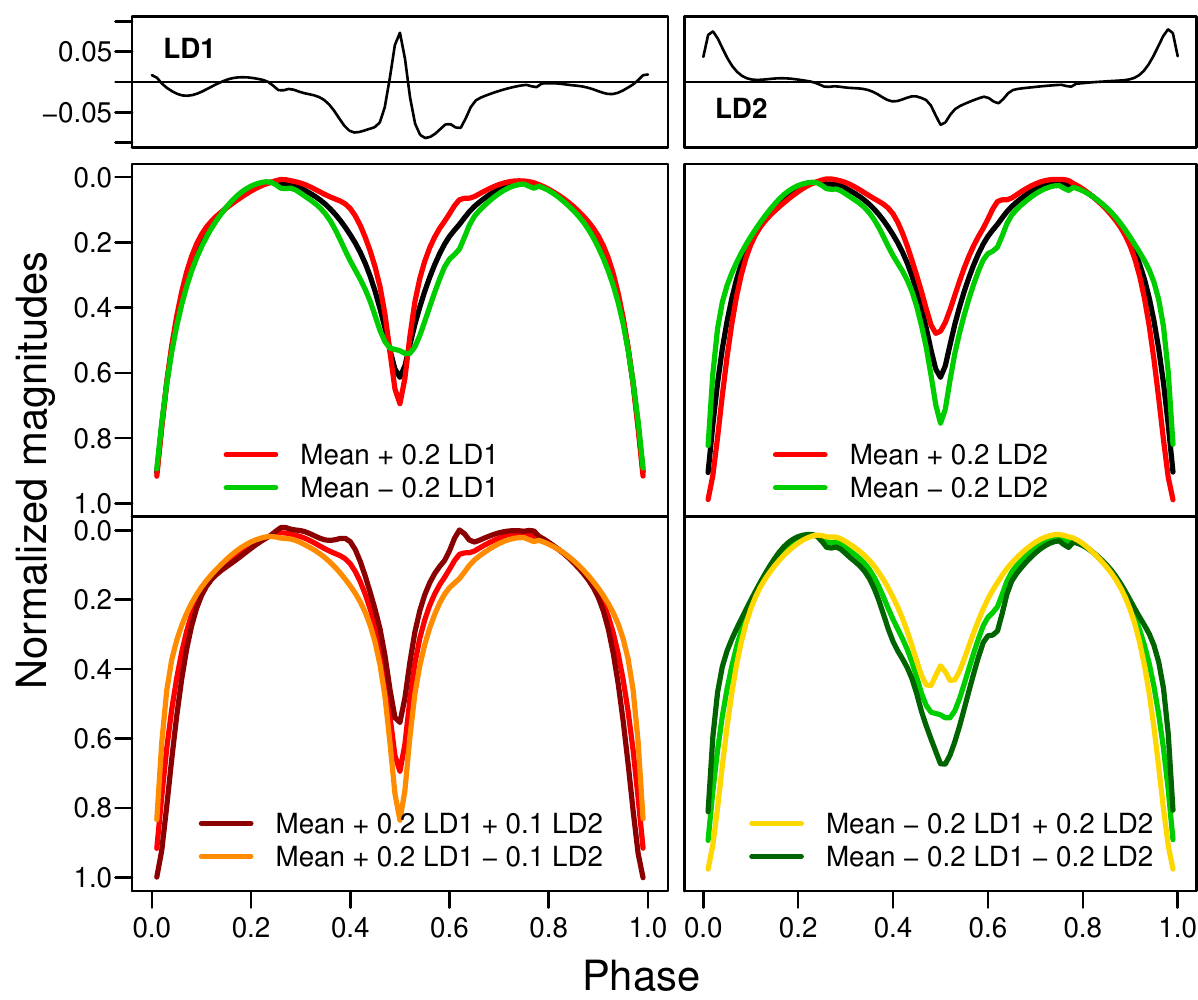}
\caption{Linear discriminant functions from the system morphology classification. For explanations, see section \ref{subsec:ld1}.}
\label{fig:LDfunctions_DSDOC}
\end{figure}

\begin{figure*}[!h]
\centering
\includegraphics[width=0.315\hsize]{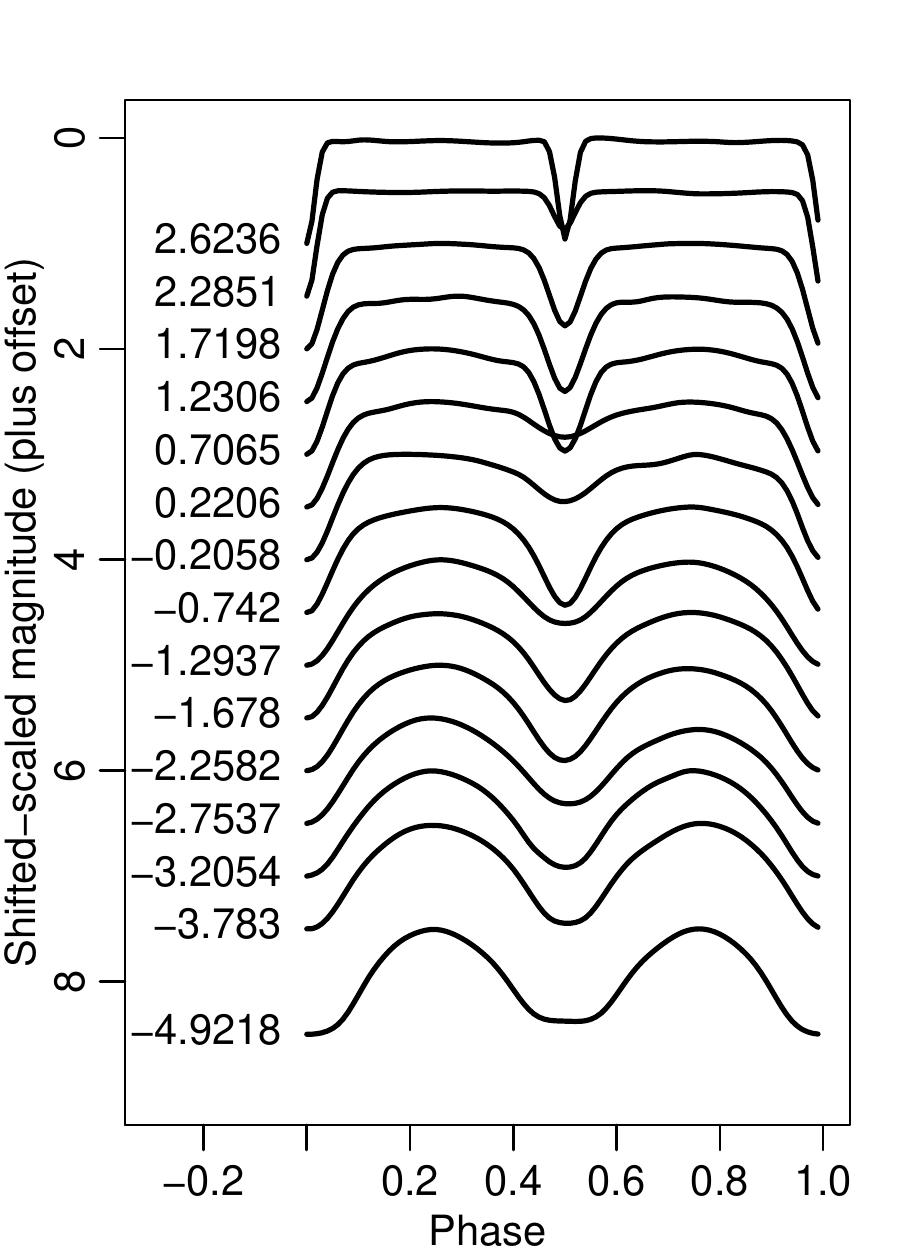}
\includegraphics[width=0.161\hsize]{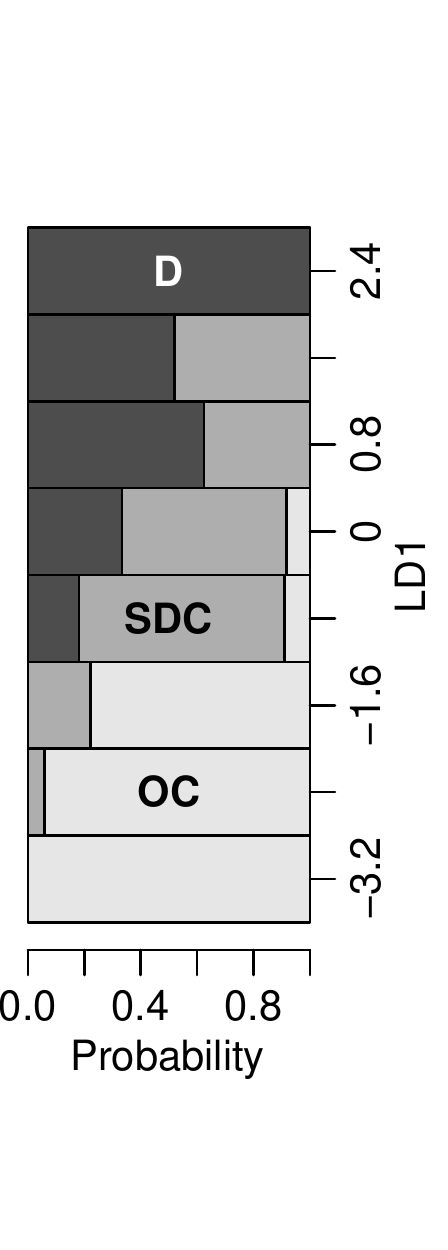}
\includegraphics[width=0.315\hsize]{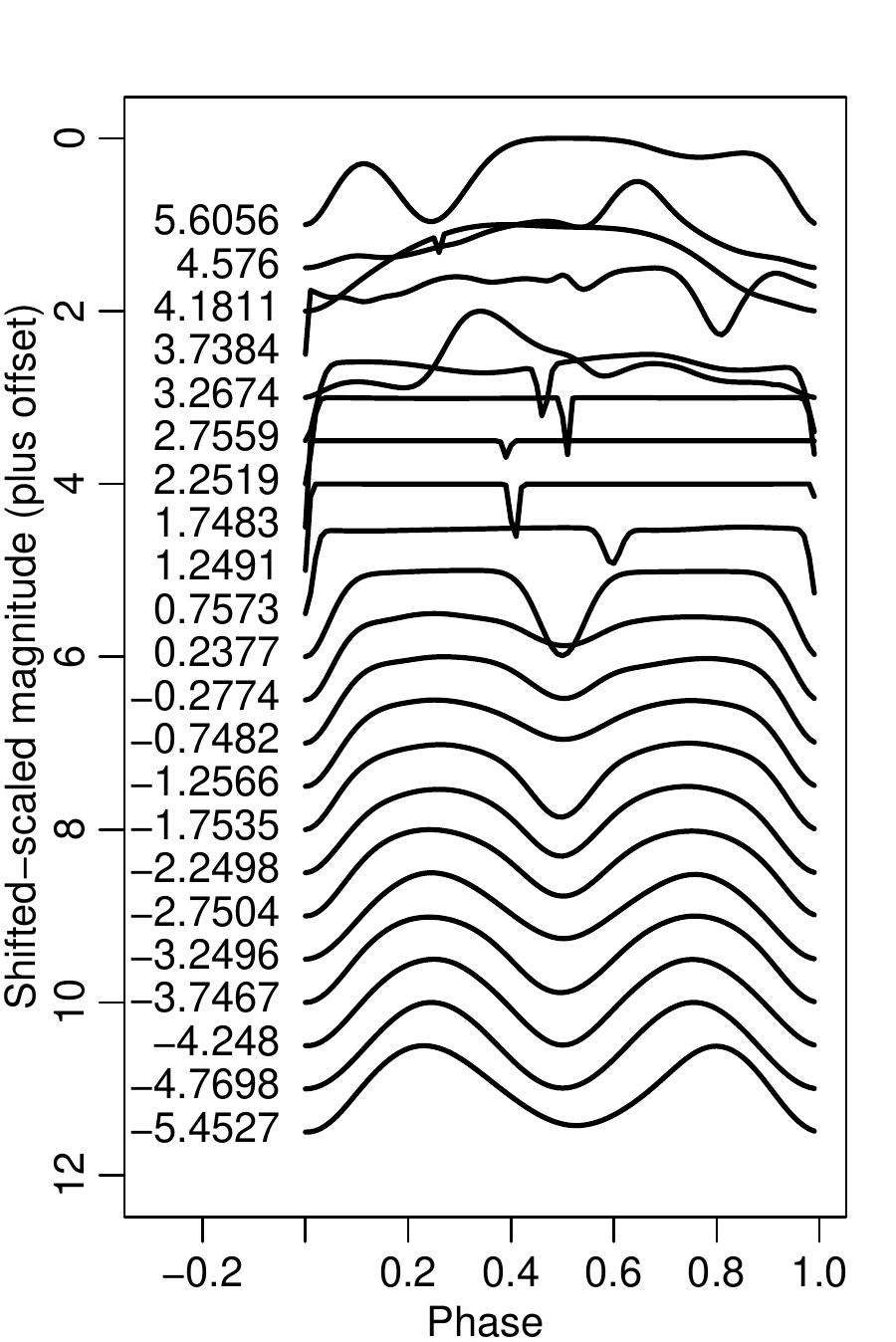}
\includegraphics[width=0.161\hsize]{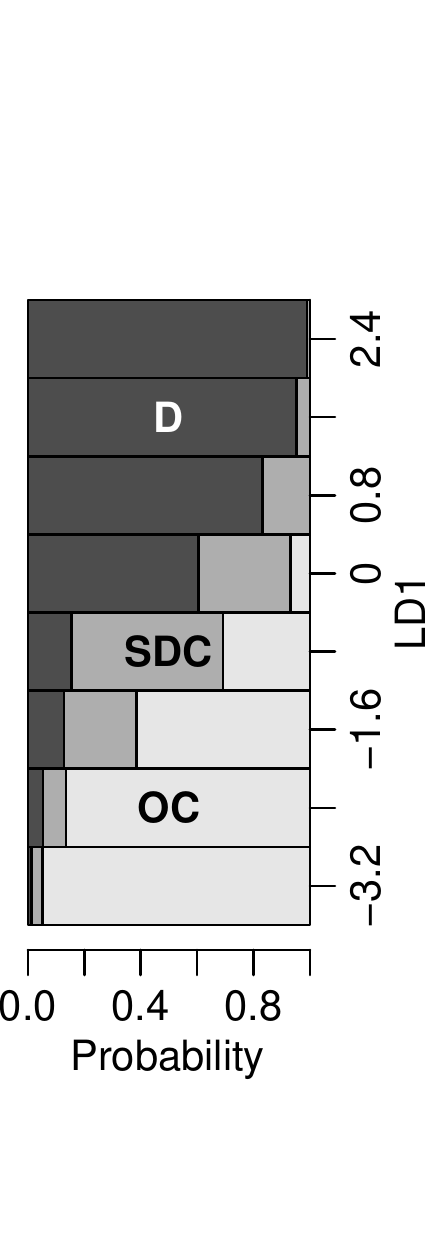}
\caption{Interpolated-scaled-shifted light curves from {\it CALEB}, ordered by their coefficients on LD1 (leftmost panel), the distribution of the system morphology classes by bins of LD1 scores (second panel from left), and the same for {\it Gaia}-sampled {\it Kepler} eclipsing binaries. In the panels showing the light curves, offsets were added for the sake of visibility. In the barplot panels, dark grey corresponds to detached systems, mid-grey, to semi-detached, near-contact and contact ones, and light grey, to overcontact ones.}
\label{fig:rainbowCALEB}
\end{figure*}

When performing an LDA only in PC space, without the period and amplitude, the linear discriminant functions are a linear combination of the PCs. They can easily be visualized similarly to the PCs as a function in phase, and given an interpretation in terms of their effect on the light curves. For a little more in-depth study of its insight into the data, we picked one specific training set, namely the same which was employed also to construct the principal component basis, and performed LDA using the first seven PCs.

Figure~\ref{fig:LDfunctions_DSDOC} shows the linear discriminant functions and their effect on the mean interpolated-standardized light curve of the training set. The discriminant functions themselves are presented in the top panel. The middle panels show the effect of these discriminant functions when subtracting them from the mean or adding them to it. Roughly speaking, while the coefficient of LD1 increases from negative to positive values, the light curve mutates from a smooth overcontact-like one with round forms to one more like a detached system, with breakpoints and sharp features. LD2 mainly adjusts the depth of the secondary eclipse: while varying LD1 towards smoother, rounder shapes will make at the same time the secondary minima shallower, varying LD2 to obtain the same overcontact-like light curves will deepen it. The joint effects can be seen in the bottom panels. The bottom left panel shows the typical light curve distortion around the mean + 0.2LD1 (the same  red  curve as in the middle left panel) once when 0.1LD2 is subtracted and once when 0.1LD2 is added. The right panel shows the corresponding distortion around the mean $-$ 0.2 LD1 (the same green curve as in the middle \textit{left} panel). 


LD1 and LD2 thus project the data into a 2-dimensional surface which is spanned by a complex variation of the light curve from overcontact-like to detached-like. The variance of the coefficients  of projection to LD1 is three times that of LD2, so we can suppose that the dominant type of variation with respect to the mean light curve in the examined population (the 120 systems in the overlap of {\it CALEB} and {\it Hipparcos}) is proportional to LD1, and LD2 only adds a second-order adjustment. We computed the linear discriminant coefficients of the whole {\it CALEB} data set, and in the leftmost panel of Figure \ref{fig:rainbowCALEB}, we arranged the interpolated-standardized light curves of 15 systems by increasing LD1 coefficient. We can observe the continuous transition from the smooth sinusoidal overcontact-like light curves to the broken line-like detached types, although the latter appears somewhat rounded. However, at each LD1 coefficient value, there is a mixture of the physical systems in the background, as the barplot in the next (second) panel shows. This is a consequence of two facts: first, the light curves of two different systems might naturally look very similar after going through the observation process, and second, details of the observed light curve are omitted when we first pre-process it, and then keep only the first few elements of its expansion using the PC basis functions. 

Thus, LD1 carries out a similar one-dimensional ordering of the light curves as the morphology parameter of \citet{matijevicetal12}, which is based on local linear embedding. In order to compare the two procedures, we created the same plot of using the interpolated-standardized {\it Gaia}-sampled light curves of {\it Kepler} eclipsing binaries. 
A sample of 22 objects, spanning approximately uniformly the whole range of LD1 coefficients on {\it Kepler} binaries, is shown in the third panel from left of Figure~\ref{fig:rainbowCALEB}.

The overall picture is very similar to that of {\it CALEB} (leftmost panel of Figure \ref{fig:rainbowCALEB}), but we can immediately notice the consequences of the sparse time sampling of {\it Gaia}. At the highest LD1 coefficients we find in overwhelming majority bad fits, but only very rarely reasonable eclipsing binary light curves. For these systems, both eclipses were undetectable with {\it Gaia} (too few or no points in eclipse), and the double Gaussian--spline fit was driven by random noise. Since these objects with {\it Gaia} observations look effectively like constant objects, this concentration at high LD1 coefficients opens a possibility for filtering out contaminating constant objects, at least partly. However, other poor fits have stochastically similar LD1 coefficient as observable eclipsing binary light curves. These are dispersed over the whole LD1 range, and cannot be filtered in such a simple way.  Apart from these, well-fitted {\it Kepler} eclipsing binary light curves at the lower LD1 values match well in character their pairs in {\it CALEB} in the leftmost panel, even though the LD1 coefficients of {\it Kepler} binaries have a larger span than for {\it CALEB}, due to the larger number of objects and to their greater diversity. 

The rightmost panel offers a comparison between the LD1 coefficient and the subjectively estimated system morphology class. The distribution of each class in each LD1 bin is very similar to what we see in the barplot showing {\it CALEB} data (second panel from the left), which is reassuring as to the quality of both the estimated system morphology classes and to the stability of our pre-processing steps and the subsequent application of FPCA and LDA.

\begin{figure}[t]
\centering
\includegraphics[width=\hsize]{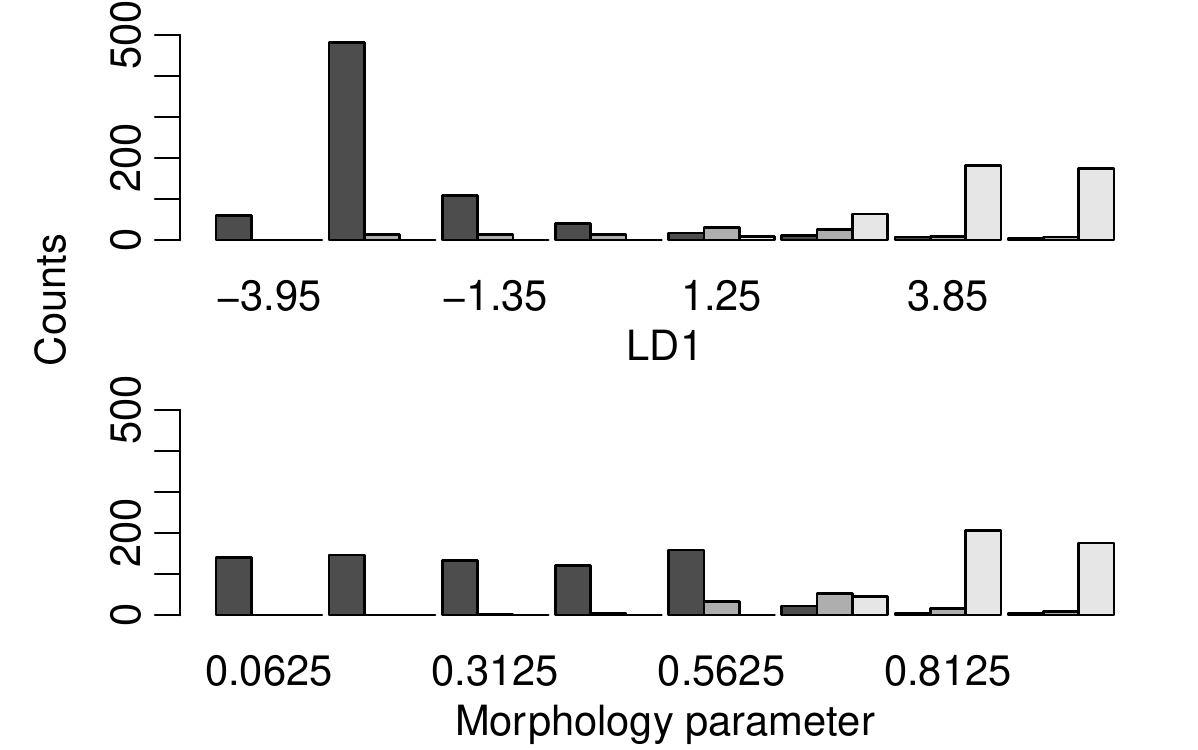}
\caption{Composition of ANN-estimated system morphology classes versus the LD1 scores (top panel) and versus the originally estimated morphology parameter (bottom panel) on the {\it Kepler} eclipsing binaries. For ease of comparison, the order of the LD1 coefficients is reversed.} 
\label{fig:KICev2_LD1_morph}
\end{figure}

\begin{figure}[t]
\centering
\includegraphics[width=\hsize]{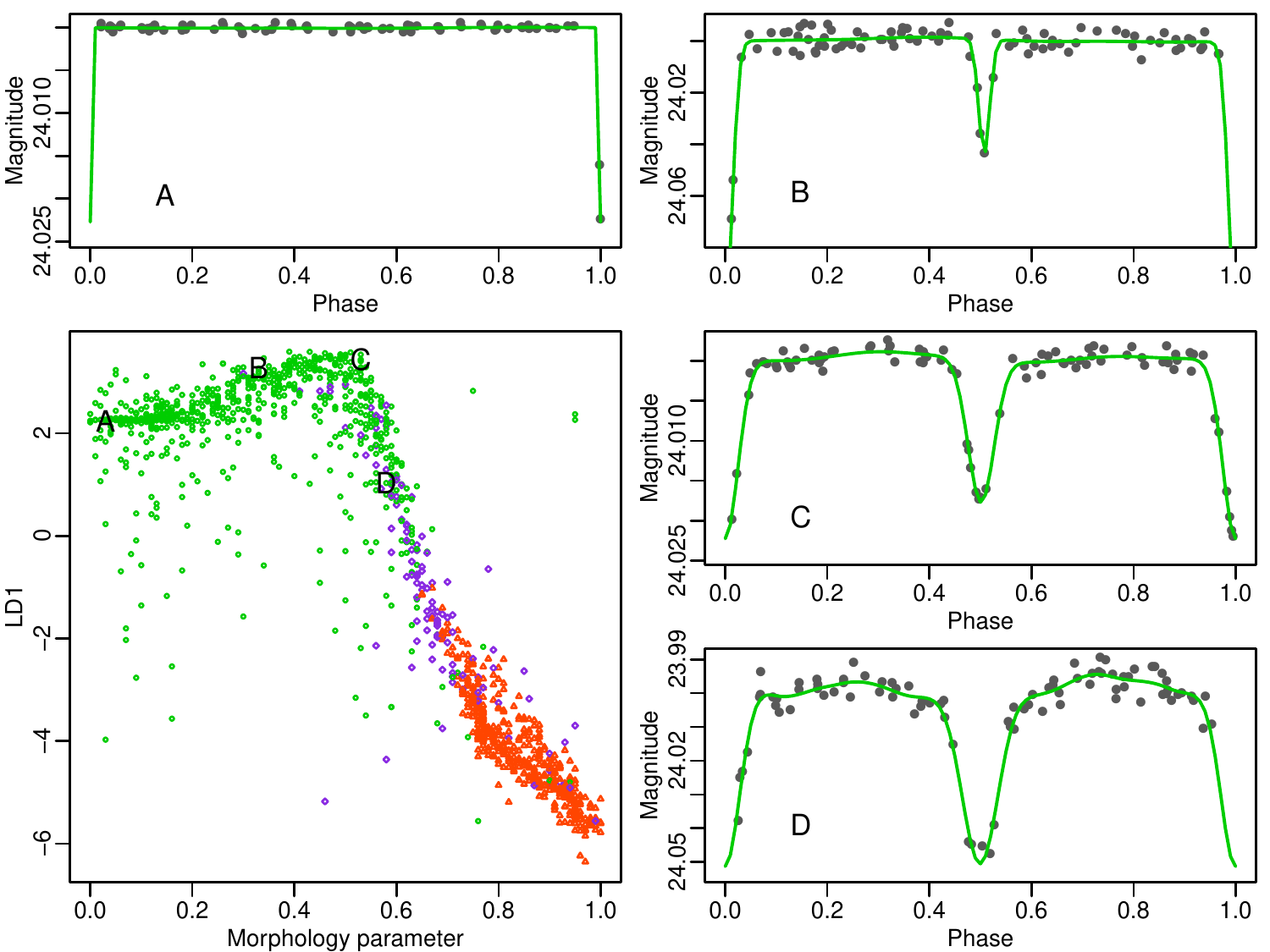}
\caption{Relation between the LD1 coefficient and the morphology parameter (large panel) and a few examples from the upper left region. Green circles: detached, violet diamonds: semi-detached (including near-contact and contact, cf. Section \ref{subsec:CALEB}), red triangles: overcontact systems. }
\label{fig:LD1LD2_vs_morph}
\end{figure}

\citet{matijevicetal12} proposes to attribute classes detached, semi-detached, overcontact (and ellipsoidal) based on the morphology parameter derived from local linear embedding. However, neither their morphology parameter on {\it Kepler}-sampled {\it Kepler} objects, nor the LD1 coefficient on {\it Gaia}-sampled {\it Kepler} objects or on the well-analysed {\it CALEB} systems leads to a cleaner separation than the classifiers above (Section \ref{subsubsec:evclass}).  A direct comparison of the class compositions at different values of the morphology parameter (derived on {\it Kepler}-sampled data\footnote{taken from \texttt{http://keplerEBs.villanova.edu}} and of the LD1 coefficient (derived on {\it Gaia}-sampled {\it Kepler} objects) is shown in Figure~\ref{fig:KICev2_LD1_morph}, in the form of grouped barplots. We see that, despite the excellent {\it Kepler} time coverage and the sparse {\it Gaia} sampling, we obtain very similar results. Though the morphology parameter stretches the detached class over a large part of the morphology parameter range and thus distinguishes variants of it, detached systems are nevertheless concentrated at one extremity of the parameter range using both methods. At the end the class occurs purely or very nearly purely. The overcontact class is gathered at the other end, again with similar purity using both methods. The semidetached class occupies the middle range by both methods, never occurring purely. There is also a slight natural mixing of even overcontact and detached systems: there are bins where all three classes occur together. 

Figure~\ref{fig:LD1LD2_vs_morph}  summarizes the direct relationship between the morphology parameter and the LD1 coefficient, on those objects from the {\it Gaia}-sampled {\it Kepler} catalogue for which the subjective class is D, SD or OC, and the morphology parameter between zero and one. The relationship is monotonic for morphology parameters between about 0.5 and 1; hence the very similar picture of class compositions in the right half of Figure~\ref{fig:KICev2_LD1_morph}. However, most of the green circles corresponding to the detached class (between morphology parameter 0 and 0.5) are much more stretched in the morphology parameter, and are in a non-monotonic relationship with LD1, although they are still concentrated at high LD1 values without too much contamination from the semi-detached systems. The reason is that local linear embedding maps the light curves to a line in a way that their density is approximately constant along the line, and so the quantity of detached systems with narrow or shallow eclipses in the {\it Kepler} data forced the method to model them in a very detailed way. LDA, which basically uses an Euclidean distance scale valid over the whole space spanned by the principal components, and corresponds to a linear path, does not sense the local geometry or the number of objects in a neighbourhood, and thus does not give more details in this region than in others.

\subsection{Unsupervised dimension reduction: SOM} \label{subsec:som}

\begin{figure*}[!h]
\centering
\includegraphics[width=0.95\hsize]{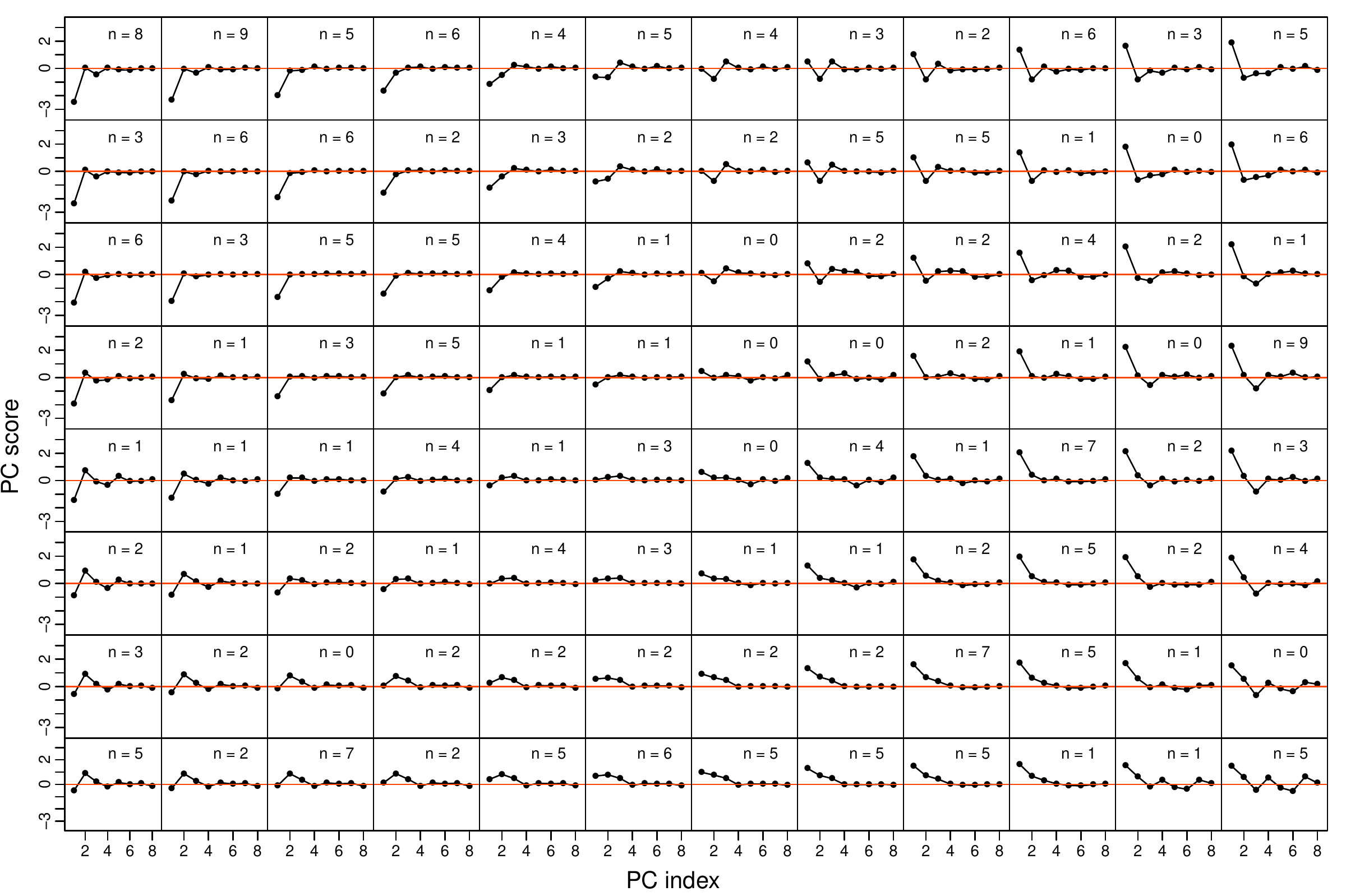}
\caption{Two-dimensional reduced space of the SOM constructed using CALEB, each cell in the plot corresponding to a gridpoint in the final fit. In each cell, we show the first eight PC decomposition coefficients PC1--PC8, determined by the gridpoint coordinates in the PC space. The numbers give the number of objects in the cell.}
\label{fig:allcalebSOM4x3rect_PC}
\end{figure*}

\begin{figure*}[!h]
\centering
\includegraphics[width=0.95\hsize]{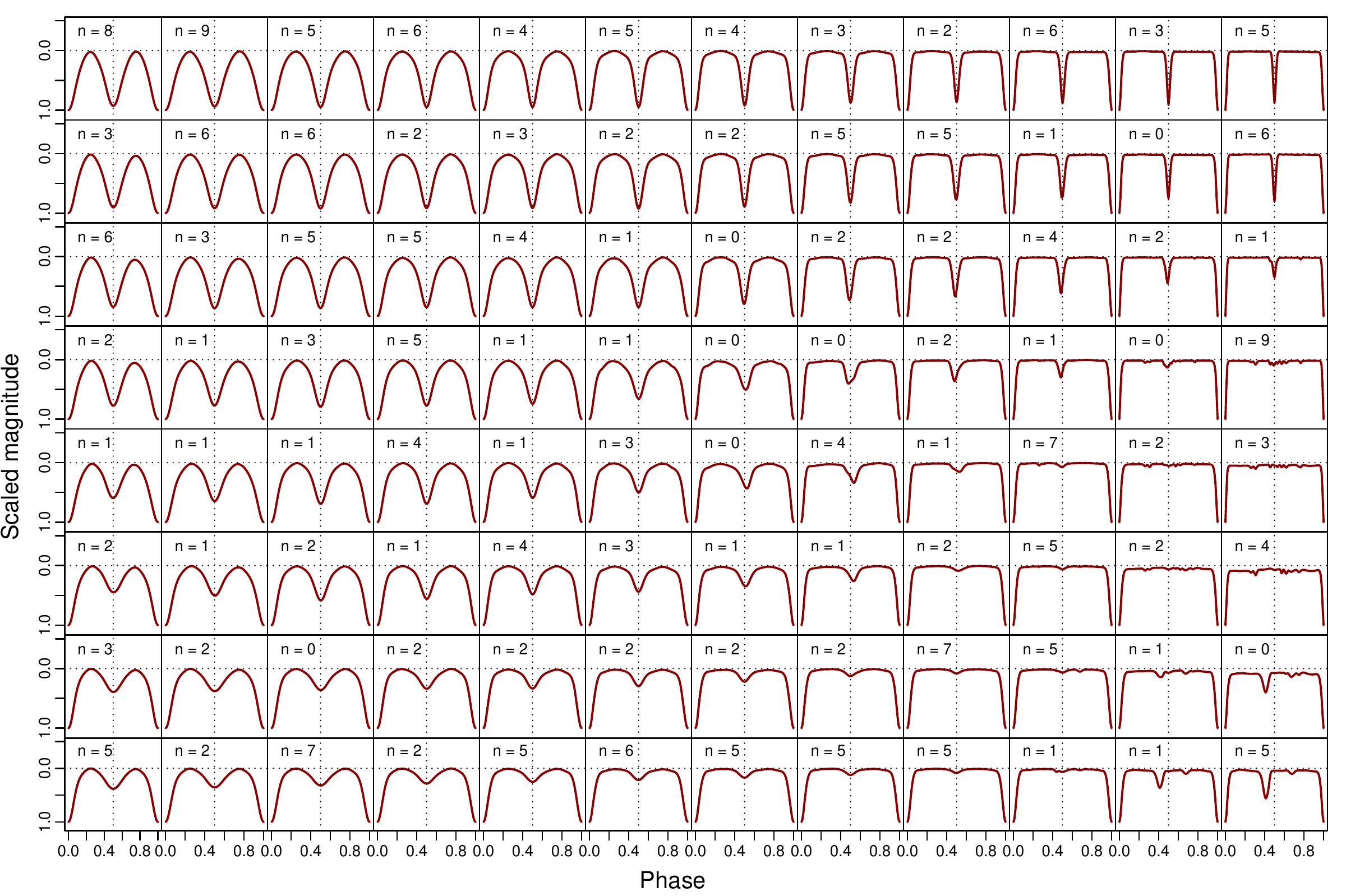}
\caption{Reconstructed light curve in each cell, corresponding to the PC profiles in Figure \ref{fig:allcalebSOM4x3rect_PC}. The numbers give the number of objects in the cell.}
\label{fig:allcalebSOM4x3rect_lc}
\end{figure*}

\begin{figure*}[!h]
\centering
\includegraphics[width=0.95\hsize]{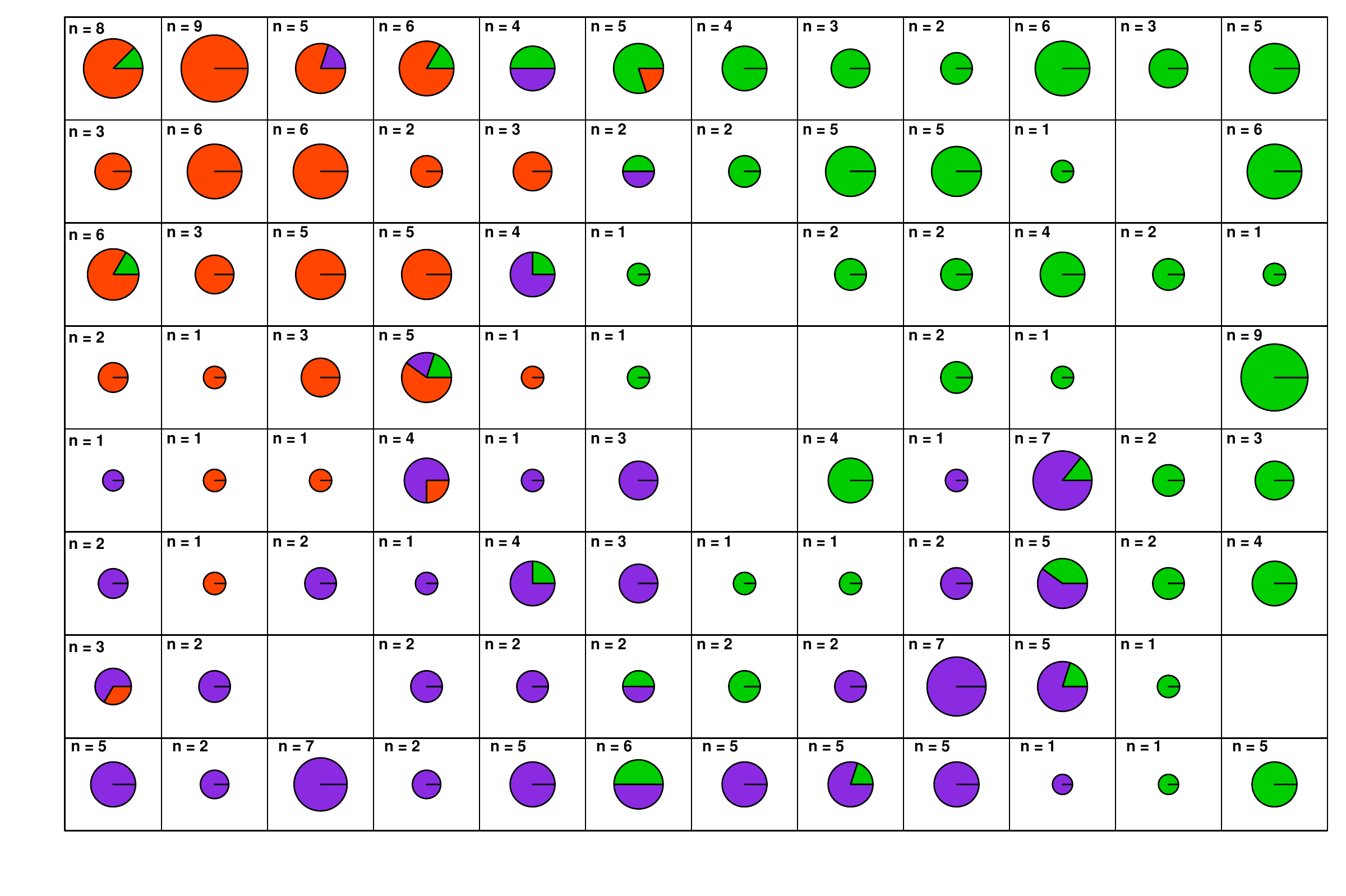}
\caption{Relative populations of the system morphology classes of {\it CALEB} in the 2-dimensional reduced space of the SOM. Red: overcontact, violet: semidetached, near-contact and contact, green: detached binaries. The area of the plotted piechart is proportional to the fraction of objects belonging to the cell in the full population.}
\label{fig:allcalebSOM4x3rect_classes}
\end{figure*}

\begin{figure*}[!h]
\centering
\includegraphics[width=0.95\hsize]{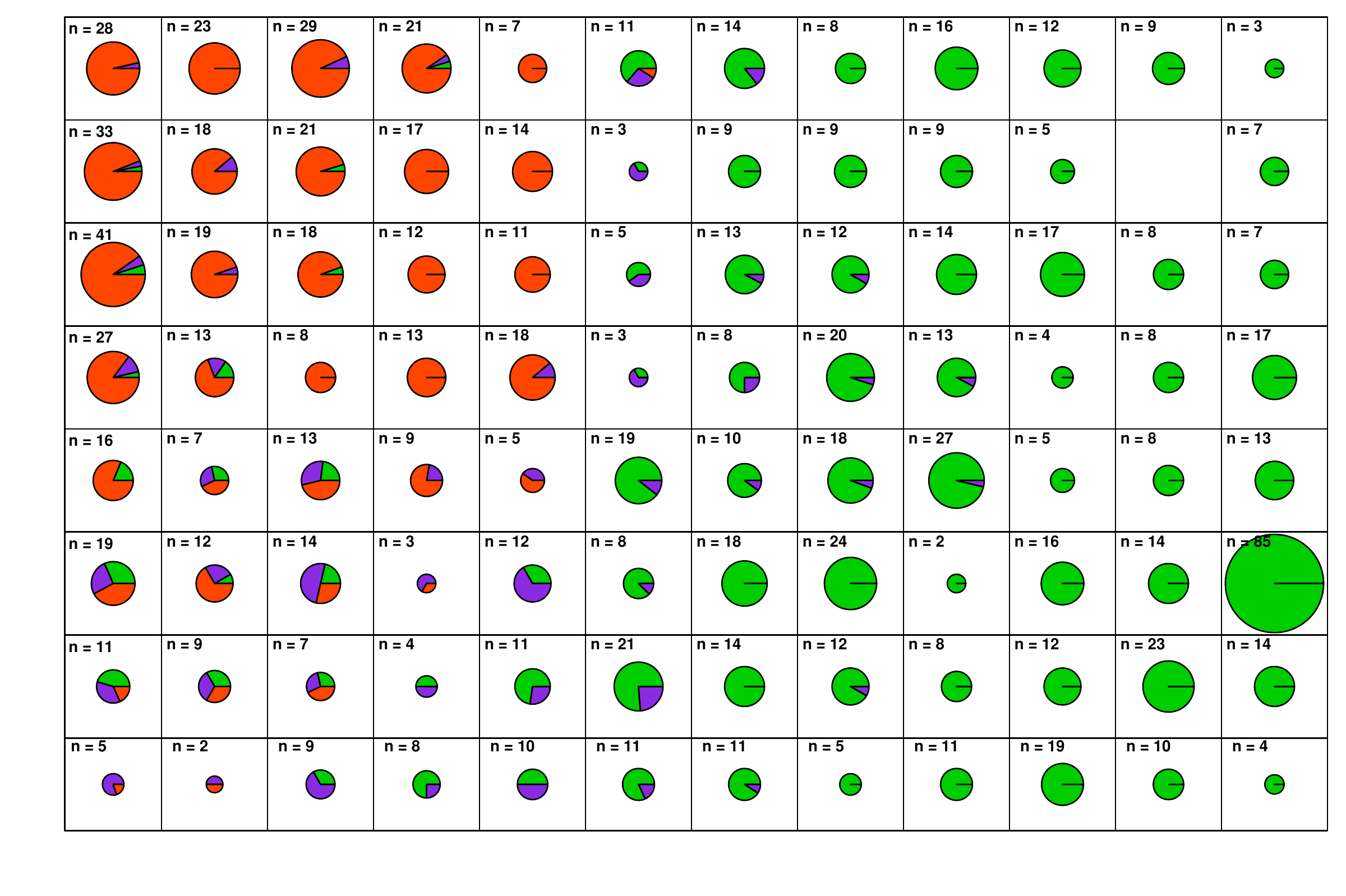}
\caption{Relative populations of the system morphology classes of {\it Kepler} systems (bottom) in the 2-dimensional reduced space of the SOM. Colour code is the same as for Figure \ref{fig:allcalebSOM4x3rect_classes}. The area of the plotted piechart is proportional to the fraction of objects belonging to the cell in the full population, except for $n<3$ in the bottom panel, which were inflated to the size of $n=3$ for visibility.}
\label{fig:kicSOM4x3rect_classes}
\end{figure*}

Principal component analysis selects directions in high-dimensional space along which the variance of the population is maximal. It is usually supposed that these directions are associated with the most informative projections of the data, and thus, PCA is considered as a fundamental dimension reduction method. However, it is inherently linear, similarly to LDA, and does not adapt automatically to the probability density peaks of the data. As a step towards methods that reflect better the complexity of the eclipsing binary light curves while at the same time decrease the dimensionality to something easier to visualize, we applied SOM. Can a more adaptive geometry discover finer features, new structures in the manifold of the eclipsing binary light curves? To enable the SOM to identify rare or weak effects such as spots, we ran it in the 99-dimensional complete PC space using the full {\it CALEB} data set.

\subsubsection{Light curves over the SOM map} \label{subsubsec:somlc}

Figures \ref{fig:allcalebSOM4x3rect_PC}--\ref{fig:kicSOM4x3rect_classes} show the resulting SOM gridpoints as a rectangular grid of cells, a ``map''. This is only for the sake of visualization: the surface traced by the grid in the space of PCs is itself not  strictly rectangular or plane-like. SOM has distorted an initial rectangular grid in a way that it passes the closest possible to the data points. The numbering of the cells (that is, the gridpoints) in the map is similar to a coordinate system, cell (1,1) is the bottom leftmost cell, and cell (12,8) is the top rightmost one. 

Figure \ref{fig:allcalebSOM4x3rect_PC} shows the values of the first eight PCs on the map. In each cell, the horizontal axis is the index of the PCs, the vertical axis shows their values (they are on a common scale). The dominant, most eye-striking changes are in PC1 and PC2. Roughly speaking, PC1 increases from left to right in each cell row, and the span of the change decreases from the upper row to the bottom row. PC2 increases from top to bottom in each column, but the range of this change is shifting downwards from left to right. There are diverse accompanying changes also in the higher-order PCs. Light curves with the most extreme and most variable values on the higher-order PCs (PC4--PC8) are concentrated in the bottom right corner.  

The light curves characterizing each cell, reconstructed from the PC1--PC99 values of the gridpoint are presented in  Figure~\ref{fig:allcalebSOM4x3rect_lc}. Smooth sinusoidal light curves are concentrated in the upper left corner; the upper panel of Figure \ref{fig:allcalebSOM4x3rect_classes}, which shows the class compositions in each cell, confirms that indeed mostly overcontact binaries can be found here. The secondary eclipse becomes narrower from left to right in all rows, and this is accompanied by the flattening of the inter-eclipse segments. Almost perfectly broken-line light curve shapes are attained in the upper right corner; this region is exclusively occupied by detached binaries, as is shown in Figure \ref{fig:allcalebSOM4x3rect_classes}. 

The most visible change in vertical direction is the overall decrease in the eclipse depths from top to bottom, as  Figure \ref{fig:allcalebSOM4x3rect_lc} shows. Correspondingly, in Figure \ref{fig:allcalebSOM4x3rect_classes}, which presents the class distributions over the SOM, we find mostly semidetached, near-contact or contact systems in the lower left corner (rounded light curves with strong tidal effects and with differing eclipse depths). Moving from left to right, ``shoulders'' and angles appear on the light curves, and in the right half, we can find broken-line light curves with  secondary eclipses getting narrower towards the right. The SOM shows a mixture of detached and semidetached systems at these places: though the lower rows are mostly occupied by the semi-detached systems, there are some detached binaries present as well. The determining character seems to be here rather the tiny or vanishing secondary eclipse depth, and not the shape of it or of the inter-eclipse phase.  

\subsubsection{Eccentricity, spots, and total eclipses in SOM} \label{subsubsec:someccspots}

The middle and lower parts of the rightmost three columns  of Figure \ref{fig:allcalebSOM4x3rect_lc} do not follow the above described almost-regular pattern. From top, the depth of the secondary eclipse decreases so quickly here that in the fourth cell from top the secondary eclipse is already barely seen. Below this, the light curves corresponding to the gridpoints seem very noisy, and exhibit not one, but a series of tiny secondary dips (recall that these are not light curves of physical eclipsing binaries, but artificial ones reconstructed from the gridpoint's coordinates in the PC space). This is so until the lowermost, rightmost couple of cells, where marked, strong eccentric dips appear on the cell light curves. These cells are regrouping mostly eccentric, in some cases single-eclipse light curves. The origin of the multiple dips on the cell light curves now becomes clear: the gridpoints corresponding to these cells, through the iterative distortions of the SOM procedure, attempt to get as close as possible to a set of noisy or eccentric binaries, whose dips can be practically anywhere on the light curve. Note that despite the fact that both eccentric and symmetric broken-line light curves (with two eclipses) need high-order PCs for modelling, SOM is apparently able to allocate separate sets of gridpoints to these groups. This suggests that although this is not apparent from any scatterplot-type visualization of the PC coefficients, there must be different higher-order structures in the PC space characterizing these groups, and the SOM is able to pull its curved surface to the proximity of the eccentric light curves.  

Unambiguous reflection effects, manifesting through the differing base level of the two eclipses, do not appear grouped in the SOM projection. The large majority of the {\it CALEB} systems we used exhibit them only relatively weakly or not at all, and so they did not force systematic distortions upon the SOM surface. Total eclipses (flat eclipse minima) are not grouped together, either. These can occur in a wide range of system configurations, with varying eclipse depths, and the SOM procedure was not able to find a concentration of them at any single typical joint PC set.

Spots can cause various distortions in the light curve. In case of a synchronized system, it is possible that the two inter-eclipses of the light curve will become unequal. Such cases do appear at least partially grouped in the SOM map, though they are relatively rare in the {\it CALEB} data (34 out of our 294 systems). Some small differences between the two maxima can be discerned in the two leftmost columns, in cells $(1,3), \ldots, (1, 7)$ and $(2, 3), \ldots, (2, 6)$: the light maximum between phases 0 and 0.5 is somewhat higher than the one between 0.5 and 1 (we will term these as ``left-sided'', by the location of the absolute brightness maximum, and the opposite case as ``right-sided''). Of the 16 spotted stars that have such left-sided asymmetry, 11 occurs in these cells, together with several others that are not fitted with spots in the {\it CALEB} data, but show the same inter-eclipse asymmetry (with two exceptions that show a slight right-sided asymmetry). The remaining spotted stars, five more with left-sided asymmetry, 12 right-sided systems, and six with equal maxima are scattered over several cells.


It appears that while the distorted surface of SOM at least partly traces the systems with a certain kind of asymmetry, it misses the cases where the asymmetry is opposite. These happen to fall into the neighbourhood of other gridpoints, where the gridpoint coordinates in the PC space do not define visibly asymmetric light curves. Recall that in order to preserve all the details of the light curves such as distortions due to eccentricity or spots, we performed the SOM in the 99-dimensional complete space, although the starting grid was the PC1--PC2 plane. In such high dimensions and with such a small sample, it may instead be due to chance that we were able to discover such similarities between systems with SOM. 

\subsubsection{{\it Gaia}-sampled {\it Kepler} binaries in SOM} \label{subsubsec:somkic}

\begin{figure}[h]
\centering
\includegraphics[width=\hsize]{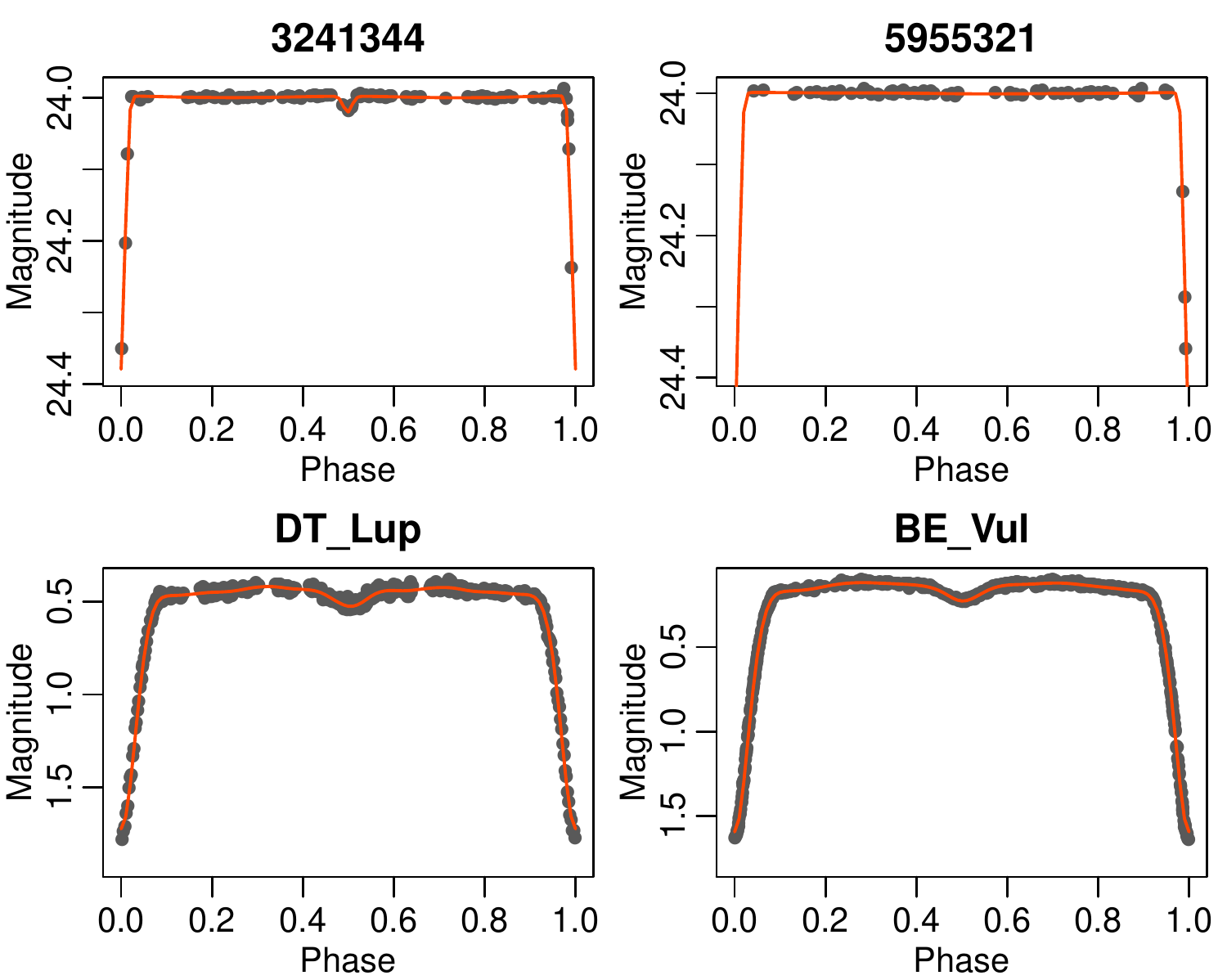}
\caption{Example light curves from cell (9,1) from {\it Kepler} (top panels) and from {\it CALEB} (bottom panels).}
\label{fig:SS_classdistr_discrep1}
\end{figure}

\begin{figure}[h]
\centering
\includegraphics[width=\hsize]{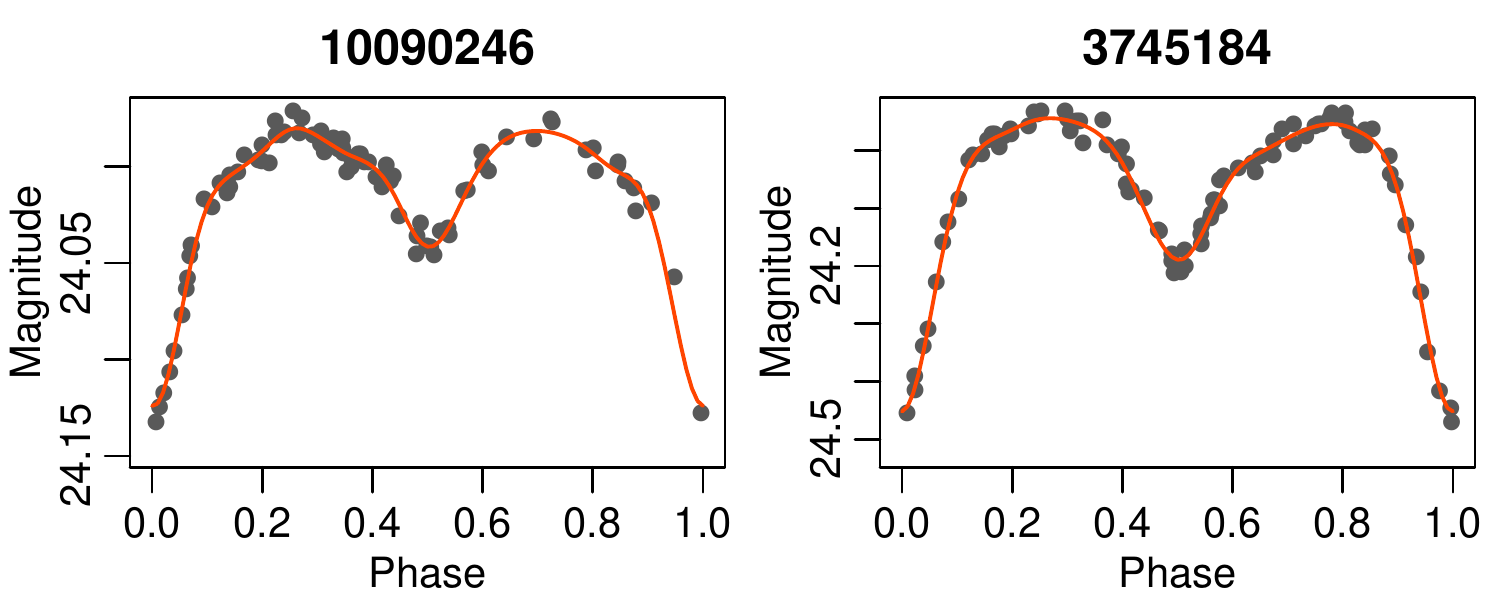}
\caption{Example light curves from cell (3,2) from {\it Kepler}. The left panel shows a system subjectively classified as D, the right panel, an OC.}
\label{fig:SS_classdistr_discrep2}
\end{figure}

\begin{figure}[h]
\centering
\includegraphics[width=\hsize]{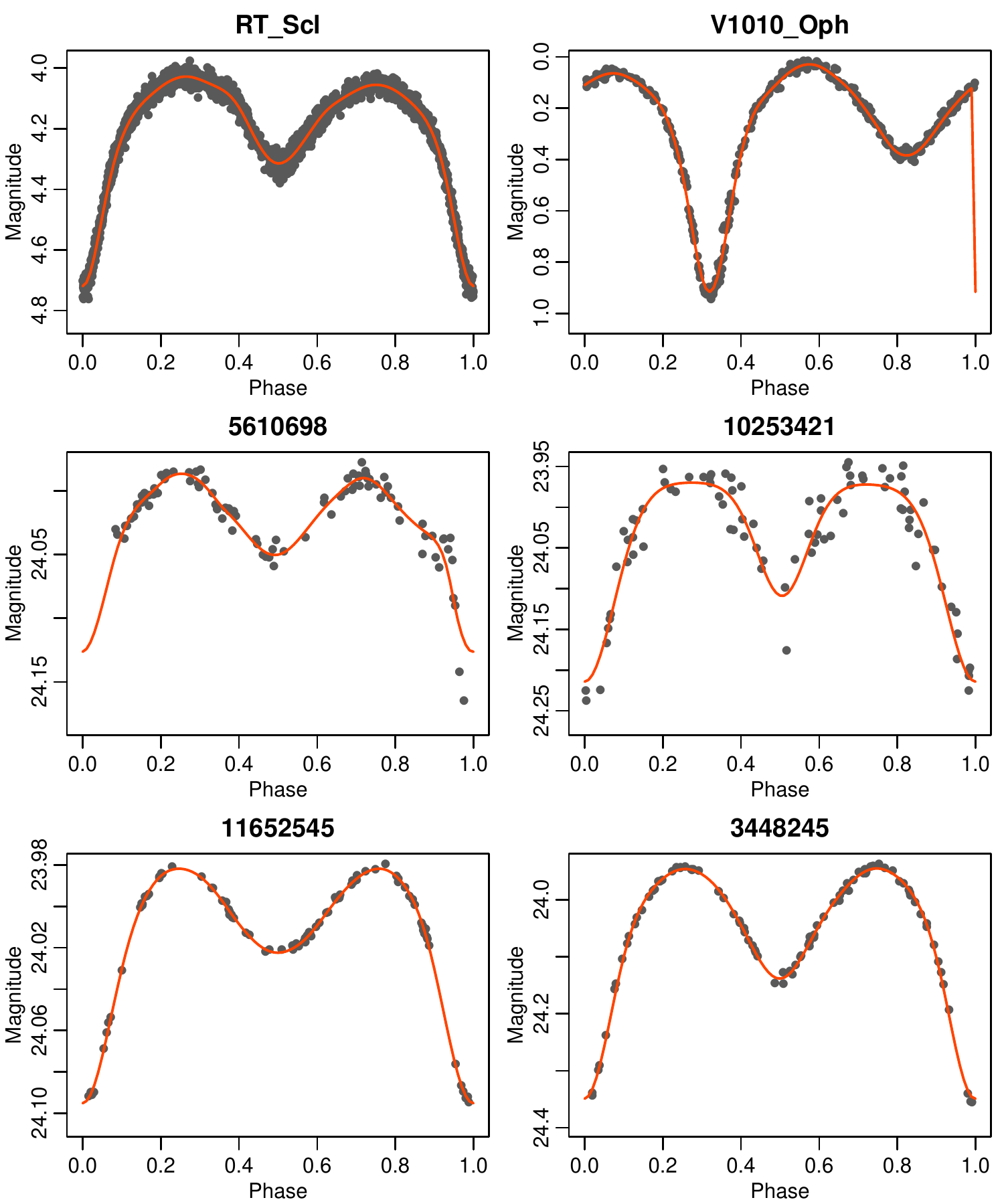}
\caption{Example light curves from cell (2,2) from {\it CALEB} (top panels, both SDC systems) and from {\it Kepler}. The middle row shows two D systems, the bottom left panel, an SD, and the bottom right, an OC. }
\label{fig:SS_classdistr_discrep3}
\end{figure}

How does this SOM model extend to a different dataset, once we saw that it gives an interpretable picture on {\it CALEB}? We can use the SOM grid trained on the well-controlled {\it CALEB} data to find the nearest neighbours of each gridpoint in any new database (with its light curves pre-processed in the same fashion and decomposed using the same FPCA basis). This informs us, on the one hand, what characteristics of the new systems can be inferred from their position on the map, and on the other hand, to assess the consequences of the systematic differences between the two surveys on the observable features of the systems. Figures \ref{fig:allcalebSOM4x3rect_classes} and  \ref{fig:kicSOM4x3rect_classes} offer a comparison between the distribution of the {\it CALEB} systems and the {\it Gaia}-sampled {\it Kepler} binaries that were classified subjectively as D, SD or OC. It shows in each cell what fraction of the complete population falls there, and within that subset, what the repartition of the system morphology classes is. 

The two maps show a fair agreement. The {\it Kepler} OC systems are concentrated in the upper left corner, D systems on the right, similarly to the {\it CALEB} map, whereas SD systems are in the lower rows, with the highest fractions in the lower left. A close look at the interpolated-shifted-scaled light curves of the {\it Gaia}-sampled {\it Kepler} binaries reveals further details.

Distributional discrepancies are evident between the detached and semidetached classes of the two surveys. The proportion of semidetached and detached classes in all cells with significant semidetached population in {\it CALEB} is very different. With {\it CALEB}, there seems to be two quite pure regions where objects classified by physical modelling as SDC are concentrated, one in the bottom left corner, and one as an intrusion into the region of detached binaries in the bottom rows, around columns 7-10. These latter cells have an overwhelming majority of detached objects in {\it Kepler}. Looking at the light curves in this region, it turns out that (in agreement with the characteristic cell light curves shown in Figure \ref{fig:allcalebSOM4x3rect_lc}) these cells contain light curves with very shallow secondary eclipses from both surveys. However, the ones from {\it CALEB} show often somewhat broader dips, smooth transition between the eclipse and the inter-eclipse part, and some weak reflection effects, while the {\it Kepler} ones have very often sharp, tiny, triangle-shaped secondary eclipses, in majority with flat inter-eclipse intervals. Figure \ref{fig:SS_classdistr_discrep1} shows two typical examples of both surveys from cell (9,1), which has a fully SDC population in {\it CALEB} (five objects), and fully D population in {\it Kepler} (11 objects). It seems that the driving effect of the SOM is the depth of the secondary eclipse, and not the light curve details around the eclipses or the inter-eclipse part. The reason for this discrepancy therefore seems to be the difference between the composition of the binaries in the {\it Kepler} data and in {\it CALEB}, which causes PCA and SOM to insufficiently model certain kind of binaries, and forces the SOM to project them into the best approximate cells.  

In general, the dominance of D systems in {\it Kepler} data in the whole right half of the SOM is due to the many systems with very narrow, very sharp or tiny eclipses. {\it Kepler} was very efficient in detecting these, thanks to its extremely dense sampling. Thus, this class is relatively more frequent among the {\it Kepler} binaries than in {\it CALEB}, where the selection was driven by completely different motivations (ultimately, the interest of scientists over decades). This is also the cause of the relatively higher fractions of the {\it Kepler} population in cells in the middle right of the map, and in those in the bottom right corner (c.f. the size of the piecharts in the cells). 

In many of these narrow-eclipse cases, the resampling by the much more sparse {\it Gaia} cadences does not obtain data in one of the minima, or have only one point. These systems end up with high probability in one of the cells characterized by a very shallow secondary. Systems with very low signal to noise ratio, or those with only one point in eclipse, may be fitted with very diverse sets of PC coefficients. Hence, these objects are scattered over the whole map; their contamination can be discovered in most of the cells, except for the upper left (overcontact) region.   

The other SDC region of {\it CALEB}, the lower left corner is also discrepant between the two surveys. Here the purely SDC population in {\it CALEB} is diluted by D and OC objects in {\it Kepler}. However, the good-quality {\it Gaia}-sampled {\it Kepler} light curves in this region are usually very similar within one cell, regardless of their manual class, and correspond well to the cell light curve. Figure \ref{fig:SS_classdistr_discrep2} shows a D and an OC system from cell (3,2), which is empty in {\it CALEB}, and contains three OC, two SD and two D systems in {\it Kepler}. In the neighbouring cell (2,2), there are two SDC systems in {\it CALEB}, and nine systems from all classes in {\it Kepler}. Figure \ref{fig:SS_classdistr_discrep3}  shows the two {\it CALEB} systems, together with two light curves classified as D (KIC 5610698 and KIC 10253421), one as SD (KIC 11652545) and one as OC (KIC 3448245) from {\it Kepler}. The figures confirm the similar aspect of the light curves in these cells, justifying their SOM position. The reasons for the mixing in this region seem to be multiple. First, the noise can smear the sharp features of D light curves (c.f. middle right panel of Figure \ref{fig:SS_classdistr_discrep3}), and the pre-processed light curves take a rounded, smoother aspect. Second, the sparse {\it Gaia} sampling has also probably lost a significant fraction of the information in the original {\it Kepler} light curve through missing fine details of the light curve, even in a less noisy case. The modelling based on {\it Gaia} observing times may therefore be rougher than the one used for the classification of {\it Kepler} light curves. Third, since these objects are not so frequent in {\it CALEB}, neither the PC decomposition, nor the SOM was adapted to their important presence. Fourth, careful, closely controlled physical modelling is more appropriate to correctly identify the morphology of the system than visual inspection or artificial intelligence. The former is subjective, depends on the individual, and can be influenced by the previously seen or simultaneously considered other light curves. The latter struggles to find the balance between oversimplification of the problem or overfitting the training set, and can be catastrophically bad in extrapolation. 

The concentration of left-sided asymmetric light curves in the middle of the first two columns persists for the {\it Gaia}-sampled {\it Kepler} data. In the cells with this feature, the majority of the {\it Kepler} light curves bears this asymmetry. Their fraction is the highest in cell (1,6), amounting to about 80\%. The rest consists of mixed symmetric and right-sided asymmetric light curves. This is so in all the asymmetric cells, albeit usually with a lower dominance. 

The lower right part of the map (roughly, the rectangle spanned by the corners (10,5) and (12,2)), contains the vast majority of eccentric binaries, similarly to {\it CALEB}. However, in the case of {\it Kepler}, they are mixed with a huge amount of systems without secondary eclipse or with only one point in eclipse. The best example of this is cell (12,3), which has an enormous population of 85 binaries. These are either eccentric or sparsely sampled and badly fitted detached systems, in an approximately 50:50 ratio.

\subsubsection{Comparison of SOM with supervised classification} \label{subsubsec:somsup}

\begin{figure*}[!h]
\centering
\includegraphics[width=0.96\hsize]{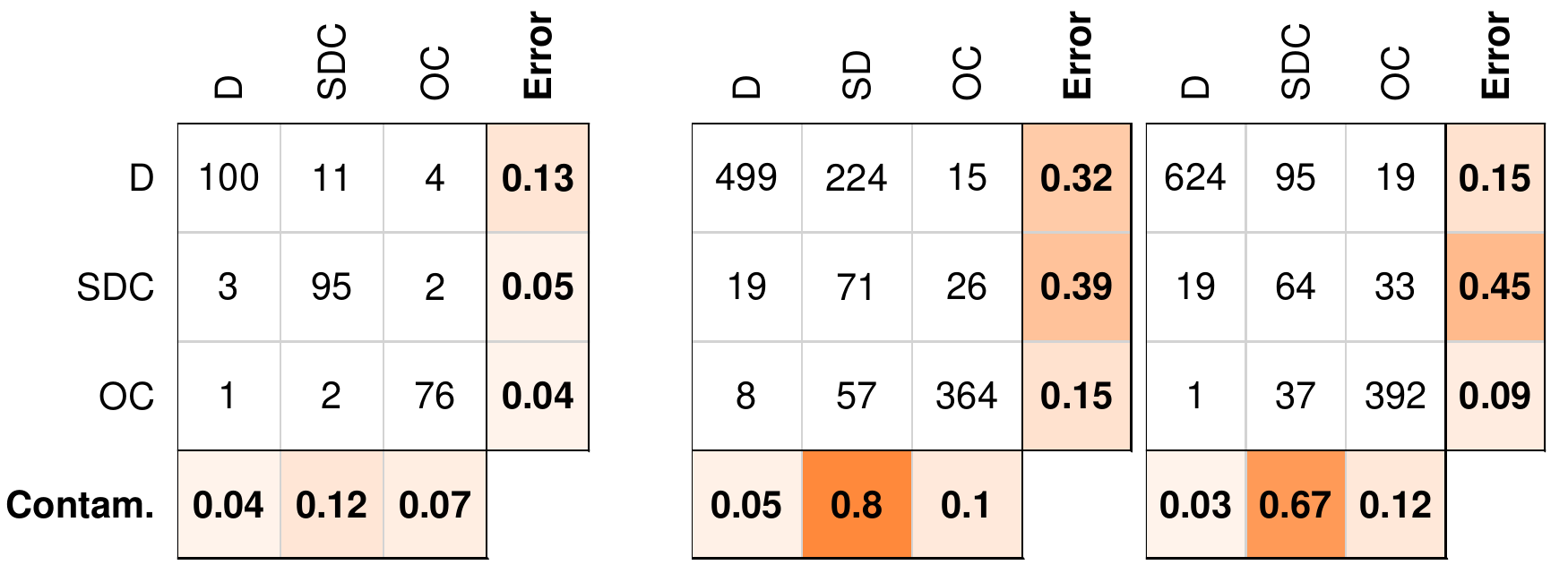}
\caption{Confusion matrices obtained by imposing the best-fit boundaries in the SOM space of \textit{CALEB}  to separate system morphology classes. Left panel: confusion matrix of  {\it CALEB} objects; middle panel: confusion matrix of estimated classes by imposing the same boundaries on the KIC objects in the SOM space; right panel: average confusion matrix obtained by applying the 500  {\it CALEB}-trained Random Forest models to  {\it Kepler} objects. The numbers in the white boxes are absolute numbers, those in the orange boxes are fractions (bottom row: contamination rate in the estimated class, rightmost column: error rate).}
\label{fig:confmatSOM_DSDOC}
\end{figure*}

Though the SOM is an unsupervised method using no input of preliminary knowledge of the class, the separation of the objects according to their system morphology on the SOM plane is remarkable. It is tempting to use it as a supervised method as well, once trained and class boundaries are determined. Attributing to each cell in the SOM surface of \textit{CALEB} (Figure \ref{fig:allcalebSOM4x3rect_classes}) its majority class, and in case of a tie, the one dominant in the neighbouring cells, we can derive the confusion matrix shown in the left panel of Figure \ref{fig:confmatSOM_DSDOC} for the \textit{CALEB} data. Comparing it to the ones produced by Random forest and LDA in Figure \ref{fig:confmat_DSDOC}, we find that the SOM-based classifier is more accurate than the others. However, we must recall that the Random forest and LDA confusion matrices are derived on test sets, data not used in the training, while the SOM uses all \textit{CALEB} data and an optimal segregation thereof. 

A better measure of the quality of a classifier is its prediction capability. The middle and right panels of Figure \ref{fig:confmatSOM_DSDOC} shed light on this. The middle panel shows the class predictions by the above segregation of SOM applied to \textit{KIC} eclipsing binaries, whereas the right panel shows results from the application of the \textit{CALEB}-trained random forest. Class labels predicted by random forest are much more accurate than those from the SOM-based classifier. The only class for which the SOM seems to slightly outperform random forest is the SDC class, as it recognizes correctly more of this class. But this is clearly due to the large area allocated to SDC in  \textit{CALEB}, and the supervised classification pays for it with an exceptionally large contamination fraction: 80\% of the objects in the predicted SDC class are in fact either D, or OC. The relative survey biases cause differences in the content of the cells, as analysed above. A cross-validation training procedure on  \textit{CALEB} to determine class boundaries might make SOM a more stable classifier, but the best use of SOM is data exploration: its ``high-resolution'' image of the topology of the distribution and its ability to describe continuous transitions between classes are extremely useful to analyse biases due to survey differences.

\section{Conclusions} \label{sec:conclusions}

In our study, we have considered statistical learning methods as tools to classify eclipsing binaries in massive surveys such as the ongoing {\it Gaia} mission and the future {\it LSST}.  We investigated the potential use of principal component analysis, random forest in classification mode, linear discriminant analysis and self-organizing maps, applied to light curves of eclipsing binaries from the {\it CALEB} and {\it Hipparcos} databases, as well as to {\it Kepler}-detected eclipsing binaries resampled according to foreseen {\it Gaia} observations. The issues touched upon are the following:~(a) efficiency of classification into two known classification systems, one according to light curve morphology (the EA/EB/EW system), the other according to the not directly observable system morphology (detached/semidetached/overcontact), and (b) data-driven discovery of low-dimensional structures in the high-dimensional manifold of the light curves. To relate the new methods and their results to already known phenomena, effects and features, and to help interpret them, we performed training and testing using well-known, thoroughly analysed objects (the {\it CALEB} database which contains the binaries together with physical modelling results, and the {\it Hipparcos} eclipsing binaries).

The starting point for these investigations is a multi-step preprocessing of the time series, using first a double Gaussian fit, then smooth splines and finally functional principal component analysis. These steps combine the advantages of all three methods: the ability of the Gaussians to provide an excellent first approximation to the potentially very high curvatures in the eclipses and the flat inter-eclipses simultaneously; the flexibility of the splines to model the residual fine details of the light curves (reflection or tidal effects, breakpoints, spots, total eclipses, various other phenomena in the light curves); and the ability of the principal component analysis to find the most important types of variations (the principal components) in the population of all light curves, and to decompose each light curve as the weighted sum of these types of variations. We found that this combination can be applied to both the relatively well-sampled, weakly noisy {\it CALEB} dataset and the {\it Kepler}-detected, but sparsely resampled binaries with a single, commonly set tuning parameter, giving stable results and generally good approximations to the real light curves. The time sampling and the different binary population in the databases do introduce some systematic differences into the classification results, since part of the {\it Kepler} population (detached binaries with very narrow or very shallow eclipses) is comparatively under-represented in the {\it CALEB} database and at the same time insufficiently sampled by {\it Gaia} (only one or two points in eclipse). Nevertheless, our combined preprocessing is still able to capture the main features of the majority of these noisy or imperfectly fitted light curves,  enable the classifiers to  recognize their nature, and put them at a reasonable position (known from {\it Kepler}) in an unsupervised structure discovery procedure.

Our main conclusion is that only two dimensions are sufficient to model a large part of the varieties in the light curves: the first two principal components account for almost 90\% of it. The dominant modes of variations can therefore be meaningfully and interpretably captured by dimension reduction methods, and the systems can roughly be characterized by only two numbers (four if we want to reach 95\%). These are sufficient to describe the relative widths and depths of the eclipses, and the presence of tidal effects by an approximate modelling of the inter-eclipses. Though higher order terms are influenced by noise, they capture some further systematic variations in the light curves, and Random Forest and linear discriminant analysis are able to give good classification accuracy into light curve morphology classes (about 95\% well-classified objects), and reasonable scores on system morphology classes (about 88-90\% well-classified objects).

Both the first linear discriminant (in one dimension) and the unsupervised SOM (in two dimensions) induce a separation of the system morphology classes, though the segregation of the classes is not clean, and represent a continuous transition rather than a classification. Each value of the first linear discriminant or each position in the SOM surface can be characterized by a vector of probabilities of the classes. Each of the three system morphology classes have a region in the reduced space where it is dominant. However, due to the gradual transitions between typical light curves of the classes, to the degeneracy implied by physically different systems producing very similar light curves, and to the time sampling that may miss small but distinctive features of the light curves, there are large regions where two or even three classes occur together. 

Beside aiding classification, the SOM trained on a controlled, well-analysed binary set offers further advantages. First, it has proven to be able to detect two more phenomena, and regroup them in specific regions of its 2-dimensional map: eccentric binaries and asymmetric maxima due most likely to spots. Second, these regions remain remarkably stable when new data, the {\it Gaia}-resampled {\it Kepler} binaries are projected to the {\it CALEB}-trained SOM surface, despite some systematic differences. Third, a careful analysis of the systematic differences yields valuable information on classification biases due to survey differences between training and new, unknown objects. The appearance of many detached systems in certain cells, the different class composition in others and the different distribution of the whole population over the SOM space helps to form a clear idea about these systematics, to rapidly understand discrepancies and unexpected features during the analysis of new data, and to interpret the results.

The caveat of the study is the small size of the training set. {\it CALEB} contains only about 300 objects with detailed physical model fits, well-known characteristics and more or less extensive information in the literature. However, random forest classification error rates improve from 12.5\% to 10.5\% when using effectively 300 training objects instead of 150. Training sets consisting of only a few hundred objects are  insufficient when the goal is the classification and analysis of millions: those volumes of data will contain samples even from the most extreme outskirts of the binary distribution and from the rarest subgroups or evolutionary states, which cannot be recognized using only those 300 objects. The creation or extension of a homogeneous, machine-readable database of reliably known binaries such as {\it CALEB} is therefore of utmost importance for the data analysis of large surveys, and reliability should optimally include also a careful, interactive modelling of the underlying physics with a binary modelling software such as the Wilson-Devinney or PHOEBE codes \citep{wilsondevinney71, prsaetal11}.

Further methodological developments will also help the application of automated analysis to the {\it Gaia} data. Our study used periods known from the databases, and therefore assessed only the performance of the methods conditional on a good knowledge of the period. An ongoing study investigates a new methodology for finding good periods for eclipsing binaries (Holl et al., in preparation). The PCA-based modelling is undirected in the sense that it adapts only to the most frequent, dominant light curve variations over the whole training population, and it does not incorporate any restriction or consistency requirement from physics on the found variation terms. Additive parametric models extending the double Gaussian fits, using specific components for tidal distortion and reflection are explored by Mowlavi et al. (submitted to A\&A). Density-based unsupervised clustering methods, yielding a different look at the natural structure of the manifold of the light curves and a new insight into the links between system morphology, physical parameters and the observed phenomenology will be considered in Kochoska et al. (under revision for A\&A). These studies will be complemented by discussions of different classification schemes and the deliverable scientific content about eclipsing binaries based on {\it Gaia} data (Mowlavi et al., in preparation, Prsa et al., in preparation).

\begin{acknowledgements}
The authors gratefully acknowledge the use of the {\it CALEB} database, created and maintained by D.~H.~Bradstreet at Eastern University, St. Davids, USA, and the discussions with the {\it Gaia} DPCG/CU7 Geneva group (L. Guy, K. Nienartowicz, D. Ord\'onez-Blanco, M. Roelens, J. Charnas, G. Jevardat) for the variability analysis of the {\it Gaia} data.
\end{acknowledgements}

\bibliographystyle{aa} 
\bibliography{bibfileOrig,bibfileAstro}

\begin{thebibliography}{53}
\expandafter\ifx\csname natexlab\endcsname\relax\def\natexlab#1{#1}\fi

\bibitem[{{Alam} {et~al.}(2015){Alam}, {Albareti}, {Allende Prieto}, {Anders},
  {Anderson}, {Anderton}, {Andrews}, {Armengaud}, {Aubourg}, {Bailey}, \&
  et~al.}]{alametal15}
{Alam}, S., {Albareti}, F.~D., {Allende Prieto}, C., {et~al.} 2015, \apjs, 219,
  12

\bibitem[{{Borucki} {et~al.}(2004){Borucki}, {Koch}, {Boss}, {Dunham},
  {Dupree}, {Geary}, {Gilliland}, {Howell}, {Jenkins}, {Kondo}, {Latham},
  {Lissauer}, \& {Reitsema}}]{boruckietal04}
{Borucki}, W., {Koch}, D., {Boss}, A., {et~al.} 2004, in ESA Special
  Publication, Vol. 538, Stellar Structure and Habitable Planet Finding, ed.
  F.~{Favata}, S.~{Aigrain}, \& A.~{Wilson}, 177--182

\bibitem[{Breiman(2001)}]{breiman01}
Breiman, L. 2001, Machine Learning, 45, 5

\bibitem[{{Carrasco Kind} \&
  {Brunner}(2014{\natexlab{a}})}]{carrascokindbrunner14a}
{Carrasco Kind}, M. \& {Brunner}, R.~J. 2014{\natexlab{a}}, \mnras, 442, 3380

\bibitem[{{Carrasco Kind} \&
  {Brunner}(2014{\natexlab{b}})}]{carrascokindbrunner14b}
{Carrasco Kind}, M. \& {Brunner}, R.~J. 2014{\natexlab{b}}, \mnras, 438, 3409

\bibitem[{{Dahlen} {et~al.}(2013){Dahlen}, {Mobasher}, {Faber}, {Ferguson},
  {Barro}, {Finkelstein}, {Finlator}, {Fontana}, {Gruetzbauch}, {Johnson},
  {Pforr}, {Salvato}, {Wiklind}, {Wuyts}, {Acquaviva}, {Dickinson}, {Guo},
  {Huang}, {Huang}, {Newman}, {Bell}, {Conselice}, {Galametz}, {Gawiser},
  {Giavalisco}, {Grogin}, {Hathi}, {Kocevski}, {Koekemoer}, {Koo}, {Lee},
  {McGrath}, {Papovich}, {Peth}, {Ryan}, {Somerville}, {Weiner}, \&
  {Wilson}}]{dahlenetal13}
{Dahlen}, T., {Mobasher}, B., {Faber}, S.~M., {et~al.} 2013, \apj, 775, 93

\bibitem[{{Deb} \& {Singh}(2009)}]{debsingh09}
{Deb}, S. \& {Singh}, H.~P. 2009, \aap, 507, 1729

\bibitem[{{Devine} {et~al.}(2016){Devine}, {Goseva-Popstojanova}, \&
  {McLaughlin}}]{devineetal16}
{Devine}, T.~R., {Goseva-Popstojanova}, K., \& {McLaughlin}, M. 2016, \mnras,
  459, 1519

\bibitem[{{Dubath} {et~al.}(2011){Dubath}, {Rimoldini}, {S{\"u}veges},
  {Blomme}, {L{\'o}pez}, {Sarro}, {De Ridder}, {Cuypers}, {Guy}, {Lecoeur},
  {Nienartowicz}, {Jan}, {Beck}, {Mowlavi}, {De Cat}, {Lebzelter}, \&
  {Eyer}}]{dubathetal11}
{Dubath}, P., {Rimoldini}, L., {S{\"u}veges}, M., {et~al.} 2011, MNRAS, 414,
  2602

\bibitem[{{Eisenstein} {et~al.}(2011){Eisenstein}, {Weinberg}, {Agol},
  {Aihara}, {Allende Prieto}, {Anderson}, {Arns}, {Aubourg}, {Bailey},
  {Balbinot}, \& et~al.}]{eisensteinetal11}
{Eisenstein}, D.~J., {Weinberg}, D.~H., {Agol}, E., {et~al.} 2011, \aj, 142, 72

\bibitem[{{European Space Agency}(1997)}]{esahipparcos}
{European Space Agency}. 1997, {The {H}ipparcos and {T}ycho catalogues} (ESA
  SP-1200)

\bibitem[{Eyer(1998)}]{eyer98}
Eyer, L. 1998, PhD thesis, Univ. of Geneva

\bibitem[{{Eyer} \& {Cuypers}(2000)}]{eyeretal00}
{Eyer}, L. \& {Cuypers}, J. 2000, in Astronomical Society of the Pacific
  Conference Series, Vol. 203, IAU Colloq. 176: The Impact of Large-Scale
  Surveys on Pulsating Star Research, ed. L.~{Szabados} \& D.~{Kurtz}, 71--72

\bibitem[{{Eyer} {et~al.}(2015){Eyer}, {Evans}, {Mowlavi}, {Lanzafame},
  {Cuypers}, {De Ridder}, {Sarro}, {Clementini}, {Guy}, {Holl}, {Ordonez},
  {Nienartowicz}, {Lecoeur-Ta{\"i}bi}, {Charnas}, {J{\'e}vardat de Fombelle},
  {Rimoldini}, \& {S{\"u}veges}}]{eyeretal15}
{Eyer}, L., {Evans}, D.~W., {Mowlavi}, N., {et~al.} 2015, IAU General Assembly,
  22, 2257301

\bibitem[{{Eyer}(2016)}]{DPACP-15}
{Eyer}, L. e.~a. 2016, \aap, in preparation

\bibitem[{{Ford}(2016)}]{ford16}
{Ford}, E.~B. 2016, Journal of Physics Conference Series, 699, 012007

\bibitem[{{Gaia~collaboration}(2016)}]{DPACP-8}
{Gaia~collaboration}. 2016, \aap

\bibitem[{{Gimenez} {et~al.}(1986){Gimenez}, {Clausen}, \&
  {Andersen}}]{gimenezetal86}
{Gimenez}, A., {Clausen}, J.~V., \& {Andersen}, J. 1986, \aap, 160, 310

\bibitem[{{Goldstein} {et~al.}(2015){Goldstein}, {D'Andrea}, {Fischer},
  {Foley}, {Gupta}, {Kessler}, {Kim}, {Nichol}, {Nugent}, {Papadopoulos},
  {Sako}, {Smith}, {Sullivan}, {Thomas}, {Wester}, {Wolf}, {Abdalla},
  {Banerji}, {Benoit-L{\'e}vy}, {Bertin}, {Brooks}, {Carnero Rosell},
  {Castander}, {da Costa}, {Covarrubias}, {DePoy}, {Desai}, {Diehl}, {Doel},
  {Eifler}, {Fausti Neto}, {Finley}, {Flaugher}, {Fosalba}, {Frieman},
  {Gerdes}, {Gruen}, {Gruendl}, {James}, {Kuehn}, {Kuropatkin}, {Lahav}, {Li},
  {Maia}, {Makler}, {March}, {Marshall}, {Martini}, {Merritt}, {Miquel},
  {Nord}, {Ogando}, {Plazas}, {Romer}, {Roodman}, {Sanchez}, {Scarpine},
  {Schubnell}, {Sevilla-Noarbe}, {Smith}, {Soares-Santos}, {Sobreira},
  {Suchyta}, {Swanson}, {Tarle}, {Thaler}, \& {Walker}}]{goldsteinetal15}
{Goldstein}, D.~A., {D'Andrea}, C.~B., {Fischer}, J.~A., {et~al.} 2015, \aj,
  150, 82

\bibitem[{Hastie {et~al.}(2009)Hastie, Tibshirani, \& Friedman}]{hastieetal}
Hastie, T., Tibshirani, R., \& Friedman, J. 2009, The Elements of Statistical
  Learning: Data Mining, Inference and Prediction, 2nd edn. (Springer-Verlag,
  New York), e-book: http://www-stat.stanford.edu/~tibs/ElemStatLearn/

\bibitem[{{Holl} {et~al.}(2014){Holl}, {Mowlavi}, {Lecoeur-Ta{\"i}bi}, \&
  {Geneva Gaia CU7 Team members}}]{holletal14}
{Holl}, B., {Mowlavi}, N., {Lecoeur-Ta{\"i}bi}, I., \& {Geneva Gaia CU7 Team
  members}. 2014, in Binary Systems, their Evolution and Environments, P2--4

\bibitem[{{Hoyle}(2016)}]{hoyle16}
{Hoyle}, B. 2016, Astronomy and Computing, 16, 34

\bibitem[{Jolliffe(2002)}]{jolliffe}
Jolliffe, I.~T. 2002, Principal Component Analysis (Springer Science \&
  Business Media, Berlin)

\bibitem[{{Kanbur} \& {Mariani}(2004)}]{kanburmariani04}
{Kanbur}, S.~M. \& {Mariani}, H. 2004, \mnras, 355, 1361

\bibitem[{{Kim} {et~al.}(2014){Kim}, {Protopapas}, {Bailer-Jones}, {Byun},
  {Chang}, {Marquette}, \& {Shin}}]{kimetal14}
{Kim}, D.-W., {Protopapas}, P., {Bailer-Jones}, C.~A.~L., {et~al.} 2014, \aap,
  566, A43

\bibitem[{{Kirk} {et~al.}(2016){Kirk}, {Conroy}, {Pr{\v s}a}, {Abdul-Masih},
  {Kochoska}, {Matijevi{\v c}}, {Hambleton}, {Barclay}, {Bloemen}, {Boyajian},
  {Doyle}, {Fulton}, {Hoekstra}, {Jek}, {Kane}, {Kostov}, {Latham}, {Mazeh},
  {Orosz}, {Pepper}, {Quarles}, {Ragozzine}, {Shporer}, {Southworth},
  {Stassun}, {Thompson}, {Welsh}, {Agol}, {Derekas}, {Devor}, {Fischer},
  {Green}, {Gropp}, {Jacobs}, {Johnston}, {LaCourse}, {Saetre}, {Schwengeler},
  {Toczyski}, {Werner}, {Garrett}, {Gore}, {Martinez}, {Spitzer}, {Stevick},
  {Thomadis}, {Vrijmoet}, {Yenawine}, {Batalha}, \& {Borucki}}]{kirk2016}
{Kirk}, B., {Conroy}, K., {Pr{\v s}a}, A., {et~al.} 2016, \aj, 151, 68

\bibitem[{Kohonen(1990)}]{kohonen90}
Kohonen, T. 1990, in Proceedings of the IEEE, Vol.~78, 1464--1479

\bibitem[{Kohonen {et~al.}(2000)Kohonen, Kaski, Lagus, Saloj{\"a}rvi, Paatero,
  \& Saarela}]{kohonenetal00}
Kohonen, T., Kaski, S., Lagus, K., {et~al.} 2000, IEEE Transactions on Neural
  Networks, 11, 574

\bibitem[{{Linnell} {et~al.}(1998){Linnell}, {Etzel}, {Hubeny}, \&
  {Olson}}]{linneletal98}
{Linnell}, A.~P., {Etzel}, P.~B., {Hubeny}, I., \& {Olson}, E.~C. 1998, \apj,
  494, 773

\bibitem[{{Mardia} {et~al.}(1979){Mardia}, {Kent}, \&
  {Bibby}}]{mardiakentbibby}
{Mardia}, K.~V., {Kent}, J.~T., \& {Bibby}, J.~M. 1979, {Multivariate analysis}

\bibitem[{{Mardirossian} {et~al.}(1980){Mardirossian}, {Mezzetti}, {Cester},
  {Giuricin}, \& {Russo}}]{mardirossianetal80}
{Mardirossian}, F., {Mezzetti}, M., {Cester}, B., {Giuricin}, G., \& {Russo},
  G. 1980, \aaps, 39, 235

\bibitem[{{Matijevi{\v c}} {et~al.}(2012){Matijevi{\v c}}, {Pr{\v s}a},
  {Orosz}, {Welsh}, {Bloemen}, \& {Barclay}}]{matijevicetal12}
{Matijevi{\v c}}, G., {Pr{\v s}a}, A., {Orosz}, J.~A., {et~al.} 2012, \aj, 143,
  123

\bibitem[{{Paltani} \& {T{\"u}rler}(2003)}]{paltaniturler03}
{Paltani}, S. \& {T{\"u}rler}, M. 2003, \apj, 583, 659

\bibitem[{{Perryman} {et~al.}(1997){Perryman}, {Lindegren}, {Kovalevsky},
  {Hoeg}, {Bastian}, {Bernacca}, {Cr{\'e}z{\'e}}, {Donati}, {Grenon},
  {Grewing}, {van Leeuwen}, {van der Marel}, {Mignard}, {Murray}, {Le Poole},
  {Schrijver}, {Turon}, {Arenou}, {Froeschl{\'e}}, \&
  {Petersen}}]{perrymanetal97}
{Perryman}, M.~A.~C., {Lindegren}, L., {Kovalevsky}, J., {et~al.} 1997, \aap,
  323

\bibitem[{{Popper}(1976)}]{popper76}
{Popper}, D.~M. 1976, \apj, 208, 142

\bibitem[{{Prsa} {et~al.}(2011){Prsa}, {Matijevic}, {Latkovic}, {Vilardell}, \&
  {Wils}}]{prsaetal11}
{Prsa}, A., {Matijevic}, G., {Latkovic}, O., {Vilardell}, F., \& {Wils}, P.
  2011, {PHOEBE: PHysics Of Eclipsing BinariEs}, Astrophysics Source Code
  Library

\bibitem[{{R Core Team}(2015)}]{R}
{R Core Team}. 2015, R: A Language and Environment for Statistical Computing, R
  Foundation for Statistical Computing, Vienna, Austria

\bibitem[{Ramsay \& Silverman(2002)}]{ramsaysilverman2}
Ramsay, J.~O. \& Silverman, B.~W. 2002, Applied functional data analysis:
  methods and case studies (Springer Science+Business Media)

\bibitem[{Ramsay \& Silverman(2010)}]{ramsaysilverman1}
Ramsay, J.~O. \& Silverman, B.~W. 2010, Functional data analysis (Springer
  Science+Business Media)

\bibitem[{{Richards} {et~al.}(2011){Richards}, {Starr}, {Butler}, {Bloom},
  {Brewer}, {Crellin-Quick}, {Higgins}, {Kennedy}, \&
  {Rischard}}]{richardsetal11}
{Richards}, J.~W., {Starr}, D.~L., {Butler}, N.~R., {et~al.} 2011, \apj, 733,
  10

\bibitem[{{Rimoldini} {et~al.}(2012){Rimoldini}, {Dubath}, {S{\"u}veges},
  {L{\'o}pez}, {Sarro}, {Blomme}, {De Ridder}, {Cuypers}, {Guy}, {Mowlavi},
  {Lecoeur-Ta{\"i}bi}, {Beck}, {Jan}, {Nienartowicz}, {Ord{\'o}{\~n}ez-Blanco},
  {Lebzelter}, \& {Eyer}}]{rimoldinietal12}
{Rimoldini}, L., {Dubath}, P., {S{\"u}veges}, M., {et~al.} 2012, \mnras, 427,
  2917

\bibitem[{{Savanov} \& {Strassmeier}(2008)}]{savanovstrassmeier08}
{Savanov}, I.~S. \& {Strassmeier}, K.~G. 2008, Astronomische Nachrichten, 329,
  364

\bibitem[{{Slawson} {et~al.}(2011){Slawson}, {Pr{\v s}a}, {Welsh}, {Orosz},
  {Rucker}, {Batalha}, {Doyle}, {Engle}, {Conroy}, {Coughlin}, {Gregg},
  {Fetherolf}, {Short}, {Windmiller}, {Fabrycky}, {Howell}, {Jenkins}, {Uddin},
  {Mullally}, {Seader}, {Thompson}, {Sanderfer}, {Borucki}, \&
  {Koch}}]{slawsonetal11}
{Slawson}, R.~W., {Pr{\v s}a}, A., {Welsh}, W.~F., {et~al.} 2011, \aj, 142, 160

\bibitem[{{Soydugan} {et~al.}(2003){Soydugan}, {Dem{\.I}rcan}, {Akan}, \&
  {Soydugan}}]{soyduganetal03}
{Soydugan}, E., {Dem{\.I}rcan}, O., {Akan}, M.~C., \& {Soydugan}, F. 2003, \aj,
  126, 1933

\bibitem[{{S{\"u}veges} {et~al.}(2012){S{\"u}veges}, {Sesar}, {V{\'a}radi},
  {Mowlavi}, {Becker}, {Ivezi{\'c}}, {Beck}, {Nienartowicz}, {Rimoldini},
  {Dubath}, {Bartholdi}, \& {Eyer}}]{suvegesetal12b}
{S{\"u}veges}, M., {Sesar}, B., {V{\'a}radi}, M., {et~al.} 2012, \mnras, 424,
  2528

\bibitem[{{Torres} {et~al.}(2010){Torres}, {Andersen}, \&
  {Gim{\'e}nez}}]{torresetal10}
{Torres}, G., {Andersen}, J., \& {Gim{\'e}nez}, A. 2010, \aapr, 18, 67

\bibitem[{{Ukwatta} {et~al.}(2016){Ukwatta}, {Wo{\'z}niak}, \&
  {Gehrels}}]{ukwattaetal16}
{Ukwatta}, T.~N., {Wo{\'z}niak}, P.~R., \& {Gehrels}, N. 2016, \mnras, 458,
  3821

\bibitem[{{Wang} \& {Lu}(1990)}]{wanglu90}
{Wang}, Y.-R. \& {Lu}, W.-X. 1990, Acta Astrophysica Sinica, 10, 389

\bibitem[{{Watson} {et~al.}(2011){Watson}, {Henden}, \& {Price}}]{watsonetal11}
{Watson}, C., {Henden}, A.~A., \& {Price}, A. 2011, VizieR Online Data Catalog,
  1

\bibitem[{{Wilson} \& {Devinney}(1971)}]{wilsondevinney71}
{Wilson}, R.~E. \& {Devinney}, E.~J. 1971, \apj, 166, 605

\bibitem[{{York} {et~al.}(2000){York}, {Adelman}, {Anderson}, {Anderson},
  {Annis}, {Bahcall}, {Bakken}, {Barkhouser}, {Bastian}, {Berman}, {Boroski},
  {Bracker}, {Briegel}, {Briggs}, {Brinkmann}, {Brunner}, {Burles}, {Carey},
  {Carr}, {Castander}, {Chen}, {Colestock}, {Connolly}, {Crocker}, {Csabai},
  {Czarapata}, {Davis}, {Doi}, {Dombeck}, {Eisenstein}, {Ellman}, {Elms},
  {Evans}, {Fan}, {Federwitz}, {Fiscelli}, {Friedman}, {Frieman}, {Fukugita},
  {Gillespie}, {Gunn}, {Gurbani}, {de Haas}, {Haldeman}, {Harris}, {Hayes},
  {Heckman}, {Hennessy}, {Hindsley}, {Holm}, {Holmgren}, {Huang}, {Hull},
  {Husby}, {Ichikawa}, {Ichikawa}, {Ivezi{\'c}}, {Kent}, {Kim}, {Kinney},
  {Klaene}, {Kleinman}, {Kleinman}, {Knapp}, {Korienek}, {Kron}, {Kunszt},
  {Lamb}, {Lee}, {Leger}, {Limmongkol}, {Lindenmeyer}, {Long}, {Loomis},
  {Loveday}, {Lucinio}, {Lupton}, {MacKinnon}, {Mannery}, {Mantsch}, {Margon},
  {McGehee}, {McKay}, {Meiksin}, {Merelli}, {Monet}, {Munn}, {Narayanan},
  {Nash}, {Neilsen}, {Neswold}, {Newberg}, {Nichol}, {Nicinski}, {Nonino},
  {Okada}, {Okamura}, {Ostriker}, {Owen}, {Pauls}, {Peoples}, {Peterson},
  {Petravick}, {Pier}, {Pope}, {Pordes}, {Prosapio}, {Rechenmacher}, {Quinn},
  {Richards}, {Richmond}, {Rivetta}, {Rockosi}, {Ruthmansdorfer}, {Sandford},
  {Schlegel}, {Schneider}, {Sekiguchi}, {Sergey}, {Shimasaku}, {Siegmund},
  {Smee}, {Smith}, {Snedden}, {Stone}, {Stoughton}, {Strauss}, {Stubbs},
  {SubbaRao}, {Szalay}, {Szapudi}, {Szokoly}, {Thakar}, {Tremonti}, {Tucker},
  {Uomoto}, {Vanden Berk}, {Vogeley}, {Waddell}, {Wang}, {Watanabe},
  {Weinberg}, {Yanny}, {Yasuda}, \& {SDSS Collaboration}}]{yorketal00}
{York}, D.~G., {Adelman}, J., {Anderson}, Jr., J.~E., {et~al.} 2000, \aj, 120,
  1579

\bibitem[{{Zevin} \& {Gravity Spy}(2016)}]{zevinspy16}
{Zevin}, M. \& {Gravity Spy}. 2016, in American Astronomical Society Meeting
  Abstracts, Vol. 228, American Astronomical Society Meeting Abstracts

\bibitem[{{Zhao} {et~al.}(2016){Zhao}, {Wang}, {Ross}, {Shandera}, {Percival},
  {Dawson}, {Kneib}, {Myers}, {Brownstein}, {Comparat}, {Delubac}, {Gao},
  {Hojjati}, {Koyama}, {McBride}, {Meza}, {Newman}, {Palanque-Delabrouille},
  {Pogosian}, {Prada}, {Rossi}, {Schneider}, {Seo}, {Tao}, {Wang}, {Y{\`e}che},
  {Zhang}, {Zhang}, {Zhou}, {Zhu}, \& {Zou}}]{zhaoetal16}
{Zhao}, G.-B., {Wang}, Y., {Ross}, A.~J., {et~al.} 2016, \mnras, 457, 2377

\end{thebibliography}

\end{document}